\documentclass[aps,prM. Vojta, R. Bulla, and W. Hofstetter, Phys. Rev. B 65,
140405 (2002).b,twocolumn,showpacs,floatfix]{revtex4-1}

\usepackage{amsmath,amssymb}
\usepackage{graphicx}% Include figure files
\usepackage{dcolumn}% Align table columns on decimal point
\usepackage{bm}% bold math
\usepackage{color}
\usepackage{comment}
\usepackage{natbib}

\newcommand{\veps}{\varepsilon}
\newcommand{\up}{\uparrow}
\newcommand{\dn}{\downarrow}

\newcommand{\pdag}{\phantom{\dag}}
\newcommand{\imp}{\text{imp}}
\renewcommand{\vec}[1]{{\boldsymbol #1}}

\begin{document}

\title{Competition between Kondo and Kitaev Physics in Kitaev clusters coupled to a fermionic bath}

\author{Tathagata Chowdhury}
\email{tatha@thp.uni-koeln.de}
\affiliation{Institut f\"{u}r Theoretische Physik, Universit\"{a}t zu K\"{o}ln,
Z\"{u}lpicher Stra$\beta$e 77a, 50937 K\"{o}ln, Germany}

\author{Achim Rosch}
\affiliation{Institut f\"{u}r Theoretische Physik, Universit\"{a}t zu K\"{o}ln,
Z\"{u}lpicher Stra$\beta$e 77a, 50937 K\"{o}ln, Germany}

\author{Ralf Bulla}
\affiliation{Institut f\"{u}r Theoretische Physik, Universit\"{a}t zu K\"{o}ln,
	Z\"{u}lpicher Stra$\beta$e 77a, 50937 K\"{o}ln, Germany}

\date{\today}

\begin{abstract}
Geometrically frustrated quantum impurities coupled to metallic leads have been shown to exhibit rich behavior with a quantum phase transition separating Kondo screened and local moment phases.
Frustration in the quantum impurity can alternatively be introduced via Kitaev-couplings between different spins of the impurity cluster.
We use the Numerical Renormalization Group (NRG) to study a range of systems where the quantum impurity comprising a Kitaev cluster is coupled to a bath of non-interacting fermions.
The models exhibits a competition between Kitaev and Kondo dominated physics depending on whether the Kitaev couplings are greater or less than the Kondo temperature.
We characterize the ground state properties of the system and determine the temperature dependence of the crossover scale for the emergence of fractionalized degrees of freedom in the model.
We also demonstrate qualitatively as well as quantitatively that in the Kondo limit, the complex impurity can be mapped to an effective two-impurity system, where the emergent spin $1/2$ comprises of both Majorana and flux degrees of freedom.
For a tetrahedral-shaped Kitaev cluster, an extra orbital degree of freedom closely related to a flux degree of freedom remains unscreened even in the presence of both Heisenberg and Kondo interactions.
\end{abstract}


\maketitle

\section{Introduction}
\label{sec:intro}

Frustration effects due to anisotropic spin interactions have been shown to have dramatic effects in quantum lattice systems.
A class of such systems consist of spin-orbit coupled Mott insulators that exhibit a strong bond-directional exchange interactions arising in spin-orbit coupled Mott insulators.
These are also known as Kitaev materials \cite{Trebst:2017}.
There is enormous interest in the physics of Kitaev materials in two as well as three dimensions as the models exhibit spin fractionalization into the elusive Majorana fermions that appear as emergent degrees of freedom coupled to a $\Bbb Z_2$ gauge field \cite{Kitaev:2006}.
The search for systems exhibiting Kitaev physics is a matter of ongoing research,
possible candidates include Na$_2$IrO$_3$ \cite{Singh:2010}, $\alpha$-Li$_2$IrO$_3$ \cite{Singh:2012} and, especially,  RuCl$_3$\cite{Plumb:2014,Nasu:2016},
where a quantized thermal Hall effect has been reported, which is interpreted as the smoking-gun signature of the Majorana edge mode of a chiral spin liquid \cite{Kasahara:2018,Vinkler-Aviv:2018,Ye:2018}.
The Kitaev honeycomb lattice is one of the rare cases, where a spin liquid can be solved exactly and it has been shown that at low temperatures, there is an ordering of the $\Bbb Z_2$ static gauge field with the spectrum being given by those of itinerant Majoranas.
However away from the Kitaev limit, it is extremely challenging to study the system using numerical as well as analytical methods.

A large group of studies have focused on the effect due to defects in the Kitaev honeycomb-lattice in 2d \cite{Willans:2010, Willans:2011} as well as in Kitaev-materials in 3d \cite{Sreejith:2016}.
The signature of these defects can act as probes in the understanding of the underlying properties of the spin liquid.
Defects can be due to the presence of vacancies in the lattice or due to the coupling with magnetic impurities at one or more lattice sites.
The main result of these studies is that in a two-dimensional Kitaev-honeycomb lattice, a vacancy binds a $\mathbb Z_2$ flux at the impurity site \cite{Willans:2010, Willans:2011}.
Additionally the response to an external magnetic field has been shown to be similar to that of a local moment at the defect site with a non-trivial dependence on the field in both gapped and gapless phases.
Subsequent studies, using perturbative scaling \cite{Das:2016, Dhochak:2010} as well as more rigorous numerical treatment using the numerical renormalization group \cite{Vojta:2016}, have shown the existence of an unstable fixed point that gives rise to a first-order flux transition between the weak-coupling flux-free phase and the strong-coupling impurity-flux phase, upon tuning the coupling between the impurity and the lattice. 
Other studies have used slave-particle mean field theories to  investigate situations involving a Kondo lattice model on the honeycomb lattice with
Kitaev interactions among the local moments, giving rise to fractionalized Fermi liquid behavior and exotic superconductivity \cite{Seifert:2018,Choi:2018}.
Here, the effective hybridization of Majorana modes and conduction electrons  due to the Kondo effect imprints superconductivity onto the conduction electron system.

While the properties of the idealized Kitaev model and of certain defects in the model are well understood, there is only limited understanding of the properties of the system when the gauge degree of freedom 
becomes dynamical due to perturbations beyond the pure Kitaev model.
We propose that quantum impurity problems can provide an alternative approach to complement the current understanding of the interplay of Majorana and gauge degrees of freedom in such situations.
In this study, we study small ``Kitaev clusters" such as those shown in Figs.\ \ref{fig:model}(a) and (b).
The clusters in question are defined to be small structures made up of spin-$1/2$'s, the only criteria being that the sites are tri-coordinated with anisotropic Kitaev interactions on each link, which allows for an analytic solution of the cluster in terms of flux and Majorana degrees of freedom.
The cluster is attached at one site to a fermionic bath via a Kondo coupling.
This couples not only the Majorana states and the conduction electrons but also gives rise to a dynamics of the flux degree of freedom on neighboring plaquettes.

From a quantum impurity perspective, the problem is that of studying a complex impurity system where the frustration effect is incorporated locally via Kitaev terms.
Frustration effects due to anisotropic spin interaction have been shown to give rise to interesting phenomena such as the spin-fractionalization in quantum spin-lattice systems.
However not much is known about the effect of frustration in the context of quantum impurity systems.
So far the focus has been on geometric frustration that can arise, for example, in a setup consisting of three quantum dots coupled to metallic leads \cite{Zitko:2008, Mitchell:2013}.
A rich range of behavior involving many-body physics has been observed in both the ferromagnetic as well antiferromagnetic regimes.
Local frustration drives the system into a phase transition separating the local moment phase with degenerate ground states and a Kondo-screened Fermi liquid phase.
These results indicate the possibility of interesting physics when frustration is introduced via Kitaev terms.
In particular, there is a competition between the Kitaev physics leading to the ordering of the $\mathbb Z_2$ degrees of freedom and the effect of the Kondo-coupling that leads to a singlet formation.
We therefore study this competition in a model which can be solved in a numerically exact way. 

We use Numerical Renormalization Group (NRG) to study the class of quantum impurity problems described above \cite{Wilson:1975,KrishnamurthyI:1980,Bulla:2008}.
It has been an extremely reliable tool that allows for an essentially exact calculation of static and dynamical properties of the quantum impurity.
The different fixed points of the model can be easily read off from the structure of the many-body NRG spectrum at the end of each iteration, thus enabling us to directly observe any fractionalized impurity degree of freedom, if any.
The NRG would also allow us to study the full crossover from the high-temperature to the low-temperature ground state(s) and the different crossover temperatures can be determined from impurity thermodynamic properties such that the entropy.

The rest of the paper is organized as follows.
The model along with the Hamiltonian is formally defined in Sec.\ \ref{sec:model}.
The numerical methods used are briefly described in Sec.\ \ref{sec:methods}, that includes the NRG as well an overview of Kitaev's exact solution of the honeycomb-lattice that is also applicable to the finite Kitaev-clusters.
The main results of this study are presented in Sec.\ \ref{sec:results}.
Firstly, in Sec.\ \ref{subsubsec:cluster} we discuss some basic properties of the Kitaev-clusters by themselves including the specific heat to illustrate spin-fractionalization in the clusters.
Impurity entropy properties of both the cube as well as the tetrahedron obtained using NRG are presented in Sec.\ \ref{subsec:Imp_S} where the dependence of the crossover temperature on model parameters is determined.
In Secs.\ \ref{subsec:2Imp} and \ref{subsec:spectrum}, we show numerically and analytically respectively, that the system can be mapped to an effective two-impurity problem.
The expectation values of the plaquette fluxes for the case of the Kitaev-cube are presented in section \ref{subsec:PF}.
Further discussion of our results and implication for future studies are presented in Sec.\ \ref{sec:discussion}.

%======================================================
\section{Model}
\label{sec:model}

\begin{figure}
\centering
\includegraphics[width=\columnwidth]{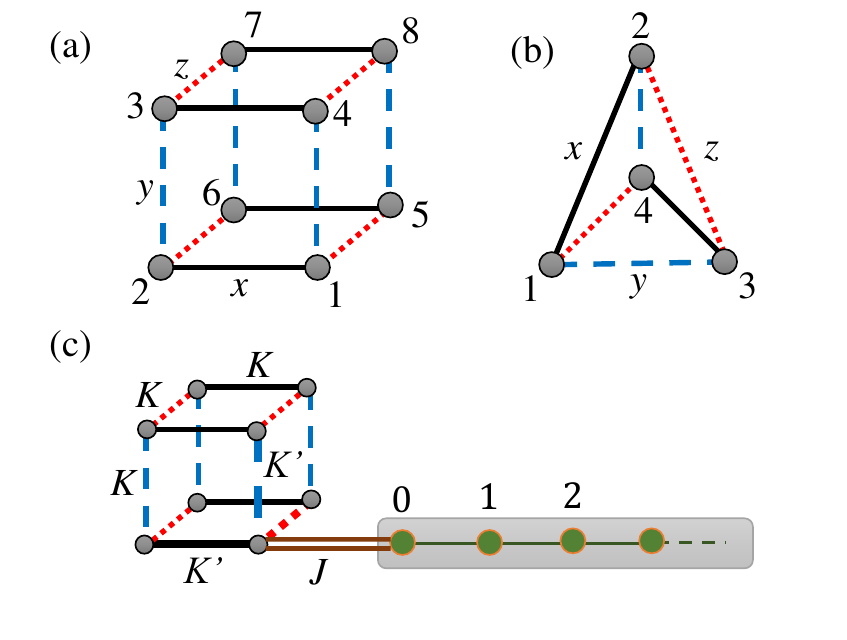}
\caption{
Finite size Kitaev clusters.
(a) A Kitaev cube consisting of eight spins.
The $x$, $y$, and $z$ interactions are shown using solid (black), dashed (blue) and dotted (red) lines respectively.
(b) A four-site Kitaev tetrahedron cluster.
(c) A schematic representation of a model under consideration.
Kitaev (cubic) cluster interacting with a bath of non-interacting fermions with Kondo coupling $J$.
The fermionic bath is attached to only one corner of the cube (site $1$).
\label{fig:model}
}
\end{figure}

In this study, we consider a range of problems where the impurity consists of a finite-size Kitaev cluster.
Each site in the cluster consists of a spin-$1/2$ that is tri-coordinated and the nature of the bonds are direction dependent like in the case of the Kitaev honeycomb lattice.
The structure of the Kitaev cluster for a cube (consisting of eight sites and twelve links) and that for a tetrahedron (consisting of four sites and six links) are shown schematically in Figs.\ \ref{fig:model}(a) and (b) respectively.
For a cluster consisting of $N_s$ spins, the sites are labeled as $1,..,N_s$ as shown in the respective figures.
In the case of the cube, we choose a configuration where the $x$, $y$, and $z$ bonds are along three different \emph{directions} and are shown using (black) solid, (blue) dashed and (red) dotted lines respectively
\footnote{
	The cube can easily be mapped to a two-dimensional geometry consisting of two quadrilaterals, one enclosing the other. Hence the bond \emph{directions} do not refer to three dimensional directions.}
\footnote{
	For a Kitaev-cluster consisting of eight sites, there exists several distinct configurations that consists of a tri-coordinated geometry. We choose the one particular configuration in this study.}.
In the case of the tetrahedral, there is only one possible configuration for the six bonds, where each pair of opposite bonds belong to either the $x$, $y$ or $z$ type of interactions.
The impurity part (Kitaev cluster) of the Hamiltonian can be written as:
\begin{align}
H_\imp = \sum_{\{ij\}} \frac{1}{4} K^{\{ij\},\gamma} \
 \hat{\sigma}^{\gamma}_i  \cdot \hat{\sigma}^{\gamma}_j,
 \label{eq:Himp}
\end{align}
where $\gamma=x,y$ or $z$ represents the type of interaction, and the Kitaev couplings $K^{\{ij\},\gamma}$ are dependent on the bond type $\{ij\}$.
In addition to the Kitaev-couplings, one can also include Heisenberg couplings given by $\sum_{\{ij\}} J_H \vec S_i \cdot \vec S_j$ along the bonds.
For $J_H\neq 0$, the gauge fields are no longer static, and instead become dynamical degrees of freedom.

We consider a geometry where this complex impurity is attached to one end of a non-interacting fermionic bath as shown schematically in Fig.\ \ref{fig:model}(c).
Although, the figure illustrates a system where the impurity consists of a Kitaev cube,
the following formalism remains unchanged for any Kitaev cluster in general.
The bath comprises a semi-infinite tight binding chain with nearest-neighbor hopping.
The fermionic site at one end of the chain (labeled as site 0 of the chain) is Kondo-coupled to one corner of the cluster (site 1 of the cluster).
The total Hamiltonian can be written as
\begin{align}
\begin{split}
H =&\ H_\imp +
\sum_{j,\sigma} \veps_j c^{\dag}_{j,\sigma}c^{\pdag}_{j,\sigma}\\
&+ \sum_{j,\sigma} t_j(c^{\dag}_{j,\sigma}c^{\pdag}_{j+1,\sigma} + c^{\dag}_{j+1,\sigma}c^{\pdag}_{j,\sigma}) \\
&+ J \bm{S}_{1}^\imp \cdot \bm{s}_{0}.
\label{eq:Hamiltonian}
\end{split}
\end{align}
Here, $c^{\dag}_{j,\sigma}$ creates a fermionic excitation at site $j\geq0$ with energy $\veps_j$ and spin configuration $\sigma=\up$ or $\dn$;
$t_j$ is the tight-binding hopping coefficient between sites $j$ and $j+1$;
$J$ is the Kondo exchange coupling between the spin $\bm{S}_{1}^\imp$ at site 1 of the cluster and $\bm s_0 = \sum_{\sigma,\sigma'} c^{\dag}_{0,\sigma}\ \frac{1}{2} \bm{\sigma}_{\sigma,\sigma'} c^{\pdag}_{0,\sigma'}$, the spin configuration at site $0$ of the fermionic chain.
In general, the density of states of the fermionic bath depends on the hopping parameters $t_j$.
We choose a density of states $\rho(\veps)=(1/2D)\Theta(|\veps-D|)$, that is constant for energies within a cutoff bandwidth $D$ and is zero otherwise.

All the Kitaev couplings are chosen to have a constant value, i.e., $K^{\{ij\},\gamma}=K$,
except for the three bonds that connect site 1 to the three nearest neighbor sites of the cluster,
such that $K^{\{ij\},\gamma}=K'$ if either $i$ or $j=1$.
These three bonds are shown using thick lines to distinguish them from the other bonds. 
There can be several variations of the preceding geometry that describe similar systems, which may give rise to further interesting phenomena.
This will be commented on in Sec.\ \ref{sec:discussion} of the paper.

%=================================================== 

\section{Methods}
\label{sec:methods}

\subsection{Numerical Renormalization Group}
\label{subsec:NRG}
We use Numerical Renormalization Group (NRG) to solve a class of problems as described by Eq.\ \ref{eq:Hamiltonian}.
The NRG has been used extensively in the context of quantum impurity systems.
It is also applicable in this work as we study an impurity in the form of a Kitaev-cluster that is attached to a fermionic bath.
However, compared to applications of the NRG so far, where the impurity consisted of only a few sites,
the Kitaev-cluster is far more complex and hence numerically more challenging to solve.
See Appendix \ref{appendix:NRG} for a brief discussion on some technical aspects of the NRG as applicable to this study.

\subsection{Exact Diagonalization of Kitaev Clusters}
\label{subsec:ED}

In order to understand the properties of the system at the different intermediate fixed points, one needs to understand the behavior of the Kitaev impurity by itself, i.e., in the absence of the fermionic bath.
This can be done using either of the two following methods:

\subsubsection{Spin-representation}
\label{subsubsec:ED-spin}
The impurity Hamiltonian can be set up using a $\sigma_i^z$ basis ($\mid \up\up\up...\up\rangle$, $\mid \dn\up\up...\up\rangle$, .., $\mid \dn\dn\dn...\dn\rangle$).
It is straightforward to solve the system by performing a direct diagonalization of the $2^{N_s} \times 2^{N_s} $ Hamiltonian matrix (for an overview refer to e.g., \cite{Sandvik:2010}).

\subsubsection{Majorana represenation}
\label{subsubsec:ED-majorana}
Alternatively, we can also study the cluster using a transformation from the spin representation to Majorana fermions as was first introduced by Kitaev \cite{Kitaev:2006}.
This method was originally developed to find an exact solution of the Kitaev honeycomb lattice, and has also been used to study other tri-coordinated lattice geometries with Kitaev couplings \cite{OBrien:2016}.
Since the finite clusters considered in this study satisfy the Kitaev criteria, they can also be exactly solved using this approach.
The key step in this technique is the fractionalization of each spin degree of freedom into four Majorana degrees of freedom.
Fermionic creation or annihilation operators can be usually expressed in terms of two Majorana fermions; hence this step increases each degree of freedom of the system by a factor of two.
Thus for a system consisting of $N_s$ spins, the dimension of the Hilbert space is artificially enlarged from $2^{N_s}$ to $2^{2N_s}$.
In the extended Hilbert space, each spin operator can be written in terms of four Majorana operators as
\begin{align}
\sigma^\gamma_i = i b^\gamma_i c_i,
\label{eq:spin_majorana}
\end{align}
where $\gamma=x,y,$ or $z$.
The Majorana operators $b^\gamma_i$ and $c_i$ are Hermitian and satisfy the following relations:
\begin{align}
\begin{split}
(b^\gamma_i)^2 = c_i^2 &= 1,\\
\{b_i^\beta,b_j^\gamma\} = 2\delta_{ij}\delta_{\beta\gamma}, \ 
\{c_i,c_j\} &= 2\delta_{ij}, \
\{c_i,b_j^\gamma\}=0. 
\end{split}
\label{eq:majorana_properties}
\end{align}
The Ising interactions in Eq.\ \eqref{eq:Himp} can then be re-expressed as
\begin{align}
\begin{split}
\sigma^\gamma \sigma^\gamma &= (i b^\gamma_i c_i)(i b^\gamma_j c_j) \\
&= -i (ib^\gamma_i b^\gamma_i) c_i c_j \\
&= -iu^\gamma_{ij}c_i c_j,
\end{split}
\label{eq:spin_uij}
\end{align}
where the bond operators $u^\gamma_{ij}=ib^\gamma_i b^\gamma_i$ are associated with the respective  links $<ij>$ in the cluster.
It can be shown that the operators $u_{ij}$'s are Hermitian and have eigenvalues $\pm 1$.
Thus, they behave like a $\Bbb Z_2$ gauge field.
The impurity Hamiltonian can thus be transformed to the form:
\begin{align}
\tilde H_\imp = \frac{i}{4} \sum_{ij} A_{ij}c_ic_j,
\label{eq:Himp_A}
\end{align}
where the $A_{ij}$'s are defined as
\begin{align}
\label{eq:A-matrix}
  A_{ij}=\begin{cases}
    \frac{K^\gamma_{ij}u^\gamma_{ij}}{2}, & \text{if $i$ and $j$ are connected}.\\
    0, & \text{otherwise}.
  \end{cases}
\end{align}
The operators $u_{ij}$ also commute with each other as well as with the impurity part of the Hamiltonian (Eq.\ \eqref{eq:Himp}).
Thus, the Hilbert space can be split up into common eigenspaces of the operators $u_{ij}$, where within the subspace, the operators can be replaced with their eigenvalues $\pm 1$.
This is akin to choosing a $\Bbb Z_2$ gauge field, see below for a discussion of the corresponding flux configurations.

In a given $\Bbb Z_2$ configuration, the matrix-elements of $A$ are thus fixed, and Eq.\ \eqref{eq:Himp_A} is reduced to a quadratic Hamiltonian.
Since $u_{ij}=-u_{ji}$, the matrix $A$ is real and skew symmetric.
Thus $iA$ is imaginary, Hermitian  and has real eigenvalues $-\epsilon_m, -\epsilon_{m-1},..,\epsilon_{m-1}, \epsilon_{m}$, with $N_s=2m$.
Equation\ \ref{eq:Himp_A} can then be reduced to the canonical form using a transformation
\begin{align}
(b'_1,b''_1,..,b'_m,b''_m)=(c_1,c_2,...,c_{2m-1},c_{2m})Q,
\end{align}
where the transformation matrix $Q$ is constructed such that odd (even) columns of $Q$ equals the real (respective imaginary) part of the eigenvectors of $iA$.
In the canonical form, the Hamiltonian can be written as 
\begin{align}
\tilde H_\text{canonical} = \frac{i}{2} \sum_{\lambda=1}^{m} \epsilon_\lambda b'_\lambda b''_\lambda
= \sum_{\lambda=1}^{m} \epsilon_\lambda ( a^\dag_\lambda a^{\pdag}_\lambda  - \frac{1}{2}),
\end{align}
where $b'_\lambda$, $b''_\lambda$ are the normal modes, and $a^\dag_\lambda=\frac{1}{2}(b'_\lambda-ib''_\lambda)$ are the corresponding fermionic creation operators.

%Thus, for a given gauge field configuration, the entire spectrum can be easily calculated by going over all the $2^{m}$ configurations of the occupation number $\{n_\lambda=a^{\dag}_\lambda a^{\pdag}_\lambda \}$.
%Note that the total spectrum of the system consists of $2^m\times2^L=2^{N_s/2}\times2^{3N/2}=2^{2N}$ states (treating each configuration $u_{ij}$ as a separate state).
%This is expected as each spin was represented using four Majorana fermions instead of two, thus adding two extra degrees of freedom per site.
%This manifests in the form of over-counting of the physical states as well of unphysical states that are altogether absent in the spin spectrum.
%This is because the eigenstates of the canonical Hamiltonian are not gauge invariant by default.
%To resolve this, one needs to define an appropriate projection operator that performs a symmetrization of the state over all gauge field configurations and thus projects out the unphysical states from the Majorana spectrum.}

For the small clusters considered in this work, neither of the above mentioned techniques (i.e., exact diagonalization using spin and Majorana representation) has significant numerical advantage over the other.
The Majorana approach, however, requires a projection back to the physical Hilbert space.
%{\color{red} Although, the Majorana approach requires a reprojection back into the physical Hilbert space and hence is more complicated.}
Both the above-mentioned methods have been employed to study the energy spectrum of the Kitaev clusters and to corroborate the results obtained with each other.
Additionally, together, they help us understand the properties of the system from two different viewpoints.

%========================================================
\subsection{Plaquette Fluxes}
\label{subsec:PF_def}

\begin{figure}
\centering
\includegraphics[width=0.8\columnwidth]{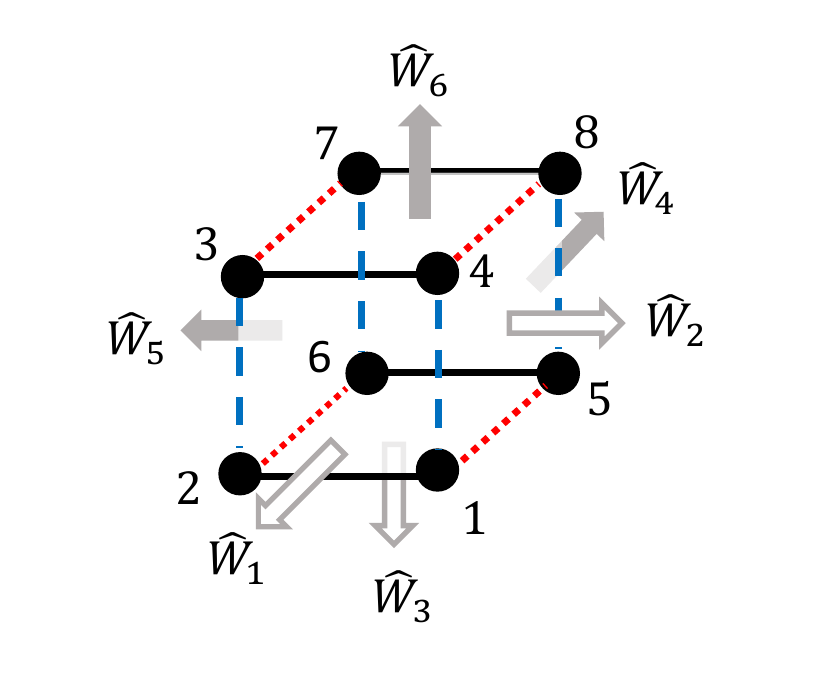}
\caption{Plaquette flux operators of the Kitaev cube.
One can define six flux operators, each associated with one of the six sides of the cube.
In the presence of a non-zero coupling between the fermionic bath and site 1 of the cube,
the operators $\hat W_{1,2,3} (\hat W_{4,5,6})$, shown using open (solid) arrows, do not commute (commute) with the Hamiltonian.
}
\label{fig:flux_config}
\end{figure}

Kitaev showed that for any closed loop in the honeycomb lattice, one can define the loop operator
\begin{align}
\hat W_l=\prod_{(ij)\in l} K_{ij},
\end{align}
in terms of the bond operators $K_{ij}=\sigma_i^\gamma \sigma_j^\gamma$.
The eigenvalues of these operators can be associated with the ``magnetic flux" through the loop.
It can be shown that for even loop length, $W_l$ has eigenvalues of $\pm1$; whereas for odd loop-lengths, it has eigenvalues of $\pm i$, and is relevant to cases where the time-reversal symmetry is broken.
The loop operators corresponding to elementary plaquettes or plaquette operators $\hat W_p$ commute with each other as well as with the Hamiltonian.
Thus, the gauge-invariant states obtained from the energy spectrum can be described in terms of the eigenvalues of the plaquette operators.
For example, in the case of the honeycomb lattice, the plaquette flux operators have a form similar to
\begin{align}
\hat W_p=\sigma_1^x \sigma_2^y \sigma_3^z \sigma_4^x \sigma_5^y \sigma_6^z,
\end{align}
where the sites $1,2,..,6$ form a hexagonal plaquette.
It is well known, that the ground state of the Kitaev honeycomb lattice corresponds to the eigenvalues $w_p=1$ for all the plaquette operators $\hat W_p$.
Elementary excitations, known as ``visons", are point-like and are associated with $w_p=-1$ or a $\pi$ flux through an elementary plaquette  \cite{Kitaev:2006,Trebst:2017}.

The above formalism can also be adapted to the case of finite clusters considered in this study.
Analogous to the plaquette operators in the Kitaev honeycomb lattice, we can define operators that correspond to the different sides of the Kitaev clusters.
For example, in the case of the cubic cluster, we can define six plaquette operators ($\hat W_p$, $p=1,2,.,.6$) corresponding to the six sides of the cube, as shown schematically in Fig.\ \ref{fig:flux_config}.
The plaquette operators (shown using arrows) are defined as
\begin{align}
\hat W_p=\prod_{(ij) \in p}\sigma_i^\gamma \sigma_j^\gamma,
\end{align}
where we have chosen a convention such that the loop is always traversed clockwise when viewed from outside.
They can also be written explicitly in terms of the spin-operators as
\begin{align}
\begin{split}
\hat  W_1 = -\sigma_1^z \sigma_2^z \sigma_3^z \sigma_4^z,
\hat W_2 = -\sigma_1^x \sigma_4^x \sigma_8^x \sigma_5^x,
\hat W_3 = -\sigma_1^y \sigma_5^y \sigma_6^y \sigma_2^y,\\
\hat  W_4 = -\sigma_5^z \sigma_8^z \sigma_7^z \sigma_6^z,
\hat W_5 = -\sigma_6^x \sigma_7^x \sigma_3^x \sigma_2^x,
\hat W_6 = -\sigma_4^y \sigma_3^y \sigma_7^y \sigma_8^y.
\end{split}
\end{align}
Since the loop length is even, the eigenvalues of $\hat W_p$ are $w_p=\pm1$,
where $w_p=+1 (-1)$ is associated with a zero ($\pi$) flux through the loop.
It should be noted that the closed geometry imposes a constraint on the flux operators such that all the operators are not independent of each other.
For example, $\hat W_1 \hat W_2 \hat W_3 \hat W_4 \hat W_5 = \hat W_6$.

In the Majorana formalism, it is straightforward to compute the eigenvalues of the flux operators.
For a given flux configuration, the plaquette flux is gauge-invariant and is given simply by the product of the bond operators:
\begin{align}
w_p = \prod_{(ij)\in p} u_{ij}.
\end{align}
In the presence of the Kondo coupling to site 1, as described in Eq.\ \eqref{eq:Hamiltonian} for $J\ne0$, the operators $\hat W_{1,2,3}$ (shown using open arrows) no longer commute with the Hamiltonian, whereas $\hat W_{4,5,6}$ (shown using solid arrows) still commute.
One can instead define an impurity flux operator $\hat W_I=\hat W_1\hat W_2\hat W_3 = \hat W_4\hat W_5 \hat W_6$, that commutes with the Hamiltonian.

Similar to the case of the cube, one can define four plaquette operators for the Kitaev-tetrahedron corresponding to the four sides of the cluster. Using the same convention used before, the plaquette operators can be expressed in terms of the spin operators as
\begin{align}
\begin{split}
	\hat W_1 = i \sigma_2^x \sigma_3^y \sigma_4^z,\ 
	\hat W_2 = i \sigma_1^x \sigma_4^y \sigma_3^z,\\
    \hat W_3 = i \sigma_4^x \sigma_1^y \sigma_2^z,\ 
    \hat W_4 = i \sigma_3^x \sigma_2^y \sigma_1^z,
\end{split}
\label{eq:tetra_flux_def}
\end{align}
Since the loop-length is odd, $\hat W_p$ have eigenvalues $w_p=\pm i$ corresponding to $\pm \pi/2$ fluxes through the elementary plaquettes. Once again, upon coupling site $1$ with a fermionic bath, the operators $\hat W_{2,3,4}$ defined in Eq.\ \eqref{eq:tetra_flux_def} no longer commute with the Hamiltonian, whereas the impurity flux operator $\hat W_I=\hat W_2 \hat W_3 \hat W_4$ commutes with the Hamiltonian.

%==============================================

\section{Results}
\label{sec:results}

In this section, we discuss the results for our study.
The main results are for systems where a Kitaev cube is coupled to fermionic bath.
However, we also include some results for that of a Kitaev tetrahedron to show
the similarities as well as the differences between the two cases.
The half-bandwidth is assumed to be $D=1$, so that it acts as a unit with respect to which all the other energy scales of the problem are measured. We adopt natural units such that $k_B = \hbar=1$.
Unless otherwise mentioned, the Kitaev couplings, $K=0.5$.
The strength of the Kitaev couplings $K'$ between site 1 and the three nearest neighbors act as a tuning parameter to study these systems.
All NRG calculations are done using a discretization parameter value of $\Lambda=3$ and keeping between 600-1000 states after each iteration.

%-----------------------------------------------------------------------------------------------------
\subsection{Kitaev Clusters}
\label{subsubsec:cluster}

\begin{figure}
	\centering
	\includegraphics[width=\columnwidth]{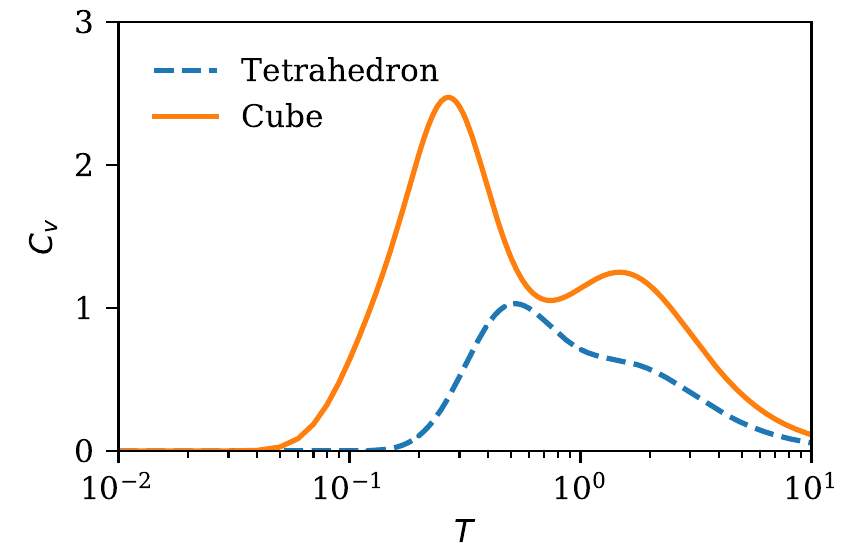}
	\caption{
		Specific heat $C_v$ of Kitaev clusters. $C_v$ is plotted against temperature $T$ on a log-scale for both a Kitaev-cube and -tetrahedron using solid and dashed lines respectively.
		The Kitaev couplings are set to $K=1$.
		Both the clusters demonstrate two peaks (less prominent in the case of the tetrahedron) as is observed for Kitaev spin systems.
			}
	\label{fig:specific_heat}
\end{figure}

Before delving into the main results, let us briefly go over some elementary properties of Kitaev clusters.
In particular, let us look at the specific heat properties of the clusters.
We perform a direct diagonalization of both Kitaev-cube and Kitaev-tetrahedron. All the Kitaev couplings are set to $K=K'=1$.
Also, the Boltzmann constant $k_B$ is assumed to be 1.
The specific heat of the systems is defined as:
\begin{align}
C_v = k_B \beta^2 (\langle E^2 \rangle - \langle E \rangle^2)
\end{align}

Figure\ \ref{fig:specific_heat} plots the specific heat $C_v$ against temperature $T$ on a logarithmic scale for both the Kitaev-cube and the Kitaev-tetrahedron using solid and dashed lines respectively.
The plot for the Kitaev-cube shows a prominent two-peak structure.
This is similar to what is observed in different Kitaev spin-lattice  systems for both $2d$ and $3d$ systems
\cite{Nasu:2014,Nasu:2015}, and is characteristic for the fractionalization of the spins into Majorana fermions and the emergent $\mathbb Z_2$ gauge fields. The high-temperature (low-temperature) peak arises from the quench of entropy carried by the Majorana fermions (the visons).
For the case of the tetrahedron, the two-peaked structure is less prominent and the two peaks are barely distinguishable due to the finite size of the cluster (the number of sites $N_s=4$).
At zero temperature, the cube has a unique ground state characterized by a $\pi$ flux through all the six plaquettes or faces. The ground state of the tetrahedron on the other hand is doubly degenerate with either $\pi/2$ or $-\pi/2$ fluxes through the four plaquettes
\footnote{It is interesting to note that the clusters have the necessary reflection symmetries such that the ground state flux configurations can be determined using Lieb's theorem. See 
E. H. Lieb, Phys. Rev. Lett. \textbf{73}, 2158 (1994).}.
This is reflected in the impurity properties of the full model as will be discussed in the next section.
Remarkably, the clusters demonstrate that similar to the case of the Kitaev spin-liquids they exhibit signatures of the fractionalization of the spins and subsequent ordering of the emergent $\mathbb{Z}_2$ gauge fields, albeit with finite size effects.
A detailed study of the energy spectrum of the clusters is presented in Sec.\ \ref{subsec:spectrum} of the paper.

%-----------------------------------------------------------------------------------------------------

\subsection{Impurity Entropy}
\label{subsec:Imp_S}

\subsubsection{Kitaev cube}

\begin{figure}
\centering
\includegraphics[width=\columnwidth]{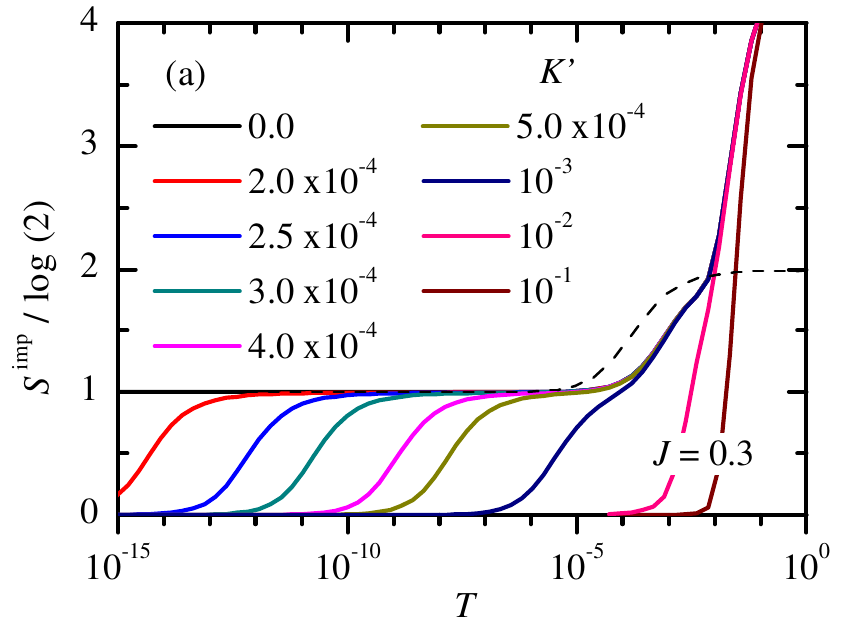} \\
\includegraphics[width=\columnwidth]{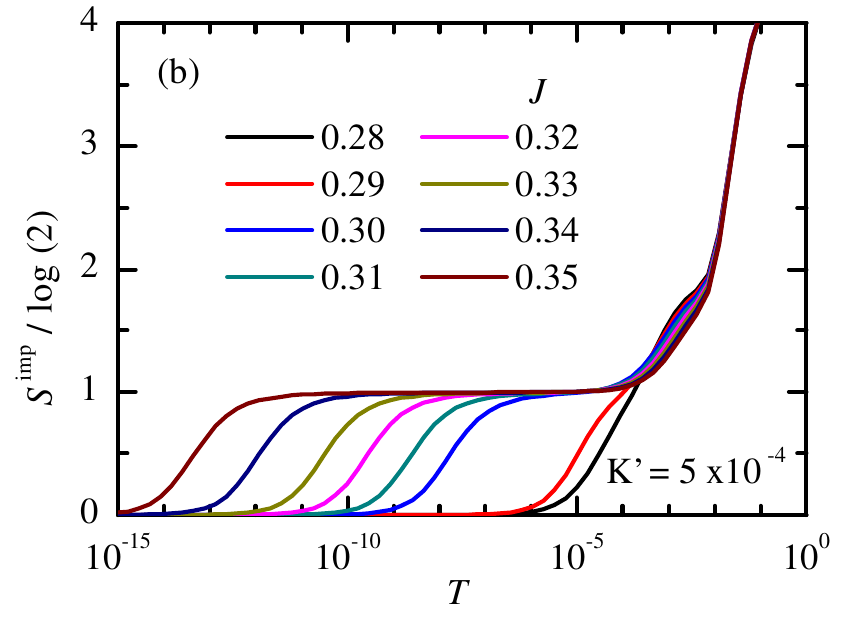}
\caption{
(a) Impurity contribution to the entropy $S^\imp$ plotted against temperature $T$ for a Kitaev cube with coupling strength $K=0.5$ for a representative case of Kondo coupling $J=0.3$.
The plots are for different values of coupling $K'$ (refer to legend).
The impurity entropy for a single impurity Kondo model with coupling $J=0.3$ is also plotted with an additional entropy of $\log 2$ using dashed lines for comparison to show the suppression of the Kondo temperature due to truncation effect.
(b) $S^\imp$ plotted against temperature $T$ for various values of the Kondo coupling $J$ keeping the Kitaev coupling $K'=5\times10^{-4}$ fixed.
}
\label{fig:cube_entropy}
\end{figure}

\begin{figure*}
\centering
\includegraphics[width=0.8\textwidth]{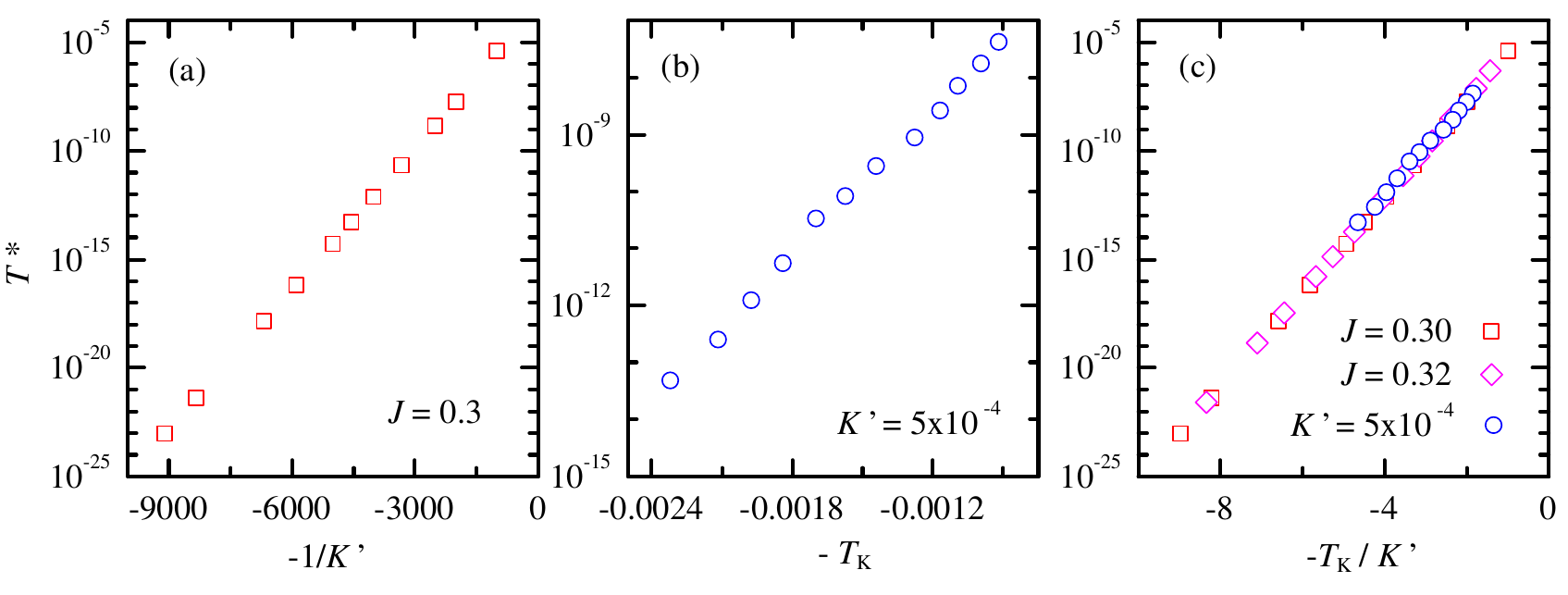}
\caption{
Crossover temperature $T^*$ for a Kitaev cube.
(a) $T^*$ plotted on a logarithmic scale against $-1/K'$ keeping $J=0.3$ constant and varying $K'$.
(same parameters as in Fig.\ \ref{fig:cube_entropy}(a)).
(b) $T^*$ plotted on a logarithmic scale against the Kondo temperature $T_K$ keeping $K'=5$x$10^{-4}$ constant and varying $J$.
(same parameters as Fig.\ \ref{fig:cube_entropy}(b)).
(c) $T^*$ plotted on a logarithmic scale against $-T_K/K'$ combining the plots shown in (a) and (b).
Also plotted are similar sets of data obtained for various values of $K'$ keeping the Kondo coupling fixed at $J=0.32$.
}
\label{fig:tstar}
\end{figure*}

\begin{figure}
\centering
\includegraphics[width=0.8\columnwidth]{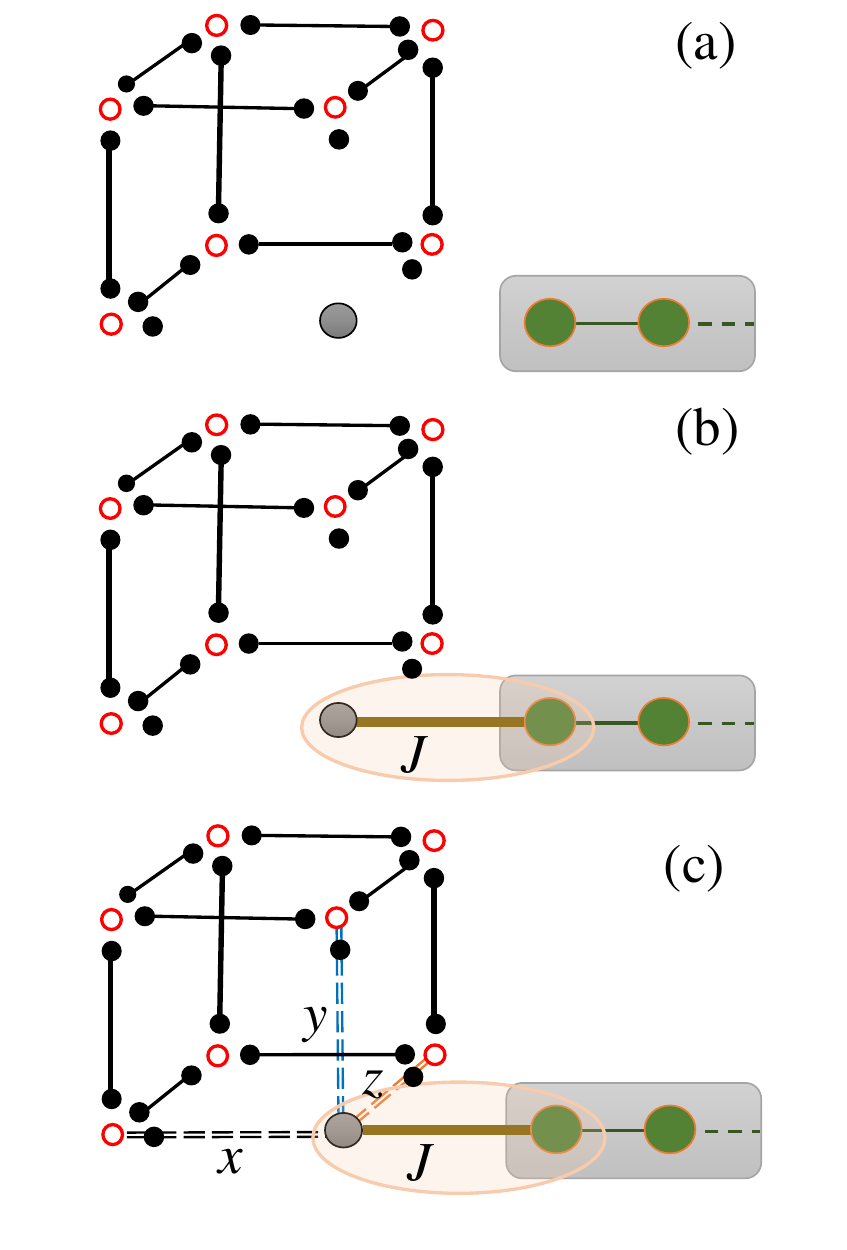}
\caption{
Intermediate fixed points of the model.
(a) $K^\prime<T_K<T$:  The system consists of a decoupled spin at site-1 and a 7-site cluster consisting of the Kitaev-cube with a vacancy.
(b) $T^*<T<T_K$: The corner site (site $1$) is Kondo-screened by the fermionic bath forming a Kondo singlet.
The rest of the cube can be represented by the bond operators $u_{i,j}$ (solid black lines) and the non-interacting $c$ Majorana fermions (red open circles) at the remaining 7 sites in addition to three dangling majoranas.
(c) $T < T^*<T_K$: The screened site interacts via Kitaev couplings $K'$ with three adjacent sites of the cube, The model can be effectively described by a Heisenberg interaction with a spin $1/2$, leading to an additional Kondo screening for temperatures $T<T^*$.
}
\label{fig:FP}
\end{figure}

Figure\ \ref{fig:cube_entropy} plots the impurity contribution to the entropy $S^\imp$ of the model for the Kitaev cube.
Subfigure (a) plots $S^\imp$ for various values of the Kitaev couplings $K'$,
the Kitaev coupling to the spin with Kondo coupling (see Fig.\ \ref{fig:model}),
for a fixed value of Kondo coupling $J=0.3$.
As the temperature decreases, the entropy initially reduces from the high temperature value of $\log 2^8$. 
This is mainly due to the sampling out of the high energy states of the spectrum of the cube by itself.
Subsequently, there is a competition between $K'$ and the Kondo temperature $T_K$.
For $K'>T_K$, the problem turns out to be a trivial one.
In this case, the impurity and the fermionic bath are decoupled from each other.
The Kitaev cube has a non-degenerate ground state with $\pi$ fluxes through all the six sides.
Hence, $S^\imp$ drops to zero at a temperature $T\sim K'$.

For $K'<T_K$, the entropy exhibits a shoulder at $S^\imp=\log 4$.
At this stage, the system can be effectively described by a 7-site cluster consisting of the Kitaev-cube minus site-1 and a decoupled free local moment at site-1 as shown schematically in Fig.\ \ref{fig:FP}(a).
The spectrum of the impurity part of the system can easily be computed using direct diagonalization by setting the strength of the Kitaev links $K'=0$.
The ground state of this cluster is found to be 4-fold degenerate and can be explained as follows.
The Kitaev-cube with a vacancy can described by a certain ground state configuration of the bond-variables $u_{ij}$ ($i,j\neq1$)  as shown using solid lines connecting the $c$ Majorana fermions (shown using black solid dots), along with three dangling majoranas at the three nearest-neighbor sites. 
This 7-site cluster has a degeneracy of 2,
and can thus be described by an effective "emergent" spin-1/2 denoted by $\vec S_e$ in the following.
Together with the spin at site 1, it forms a 4-dimensional Hilbert space with an associated entropy of $\log 4$. As we discuss in more detail in Sec.\ \ref{subsec:spectrum}, these spins encode both majorana and flux degrees of freedom. 
As the temperature is reduced, the spin at site-1 forms a Kondo singlet with the fermionic bath at the characteristic Kondo temperature $T_K \sim \exp(-1/J)$ ($T_K\approx10^{-3}$ in Fig.\ \ref{fig:cube_entropy}(b)).
Thus $S^\imp$ decreases from $\log 4$ to $\log 2$ as is demonstrated by all the curves for $K' \leq 0.001$.
In the case of the cubic cluster, we numerically determine $T_K$ as the temperature at which the impurity entropy $S^\imp$ crosses $\frac{3}{2}\log 2$ while decreasing from $\log 4$ to $\log 2$. 
As the temperature is further lowered, the Kitaev-like interactions between the corner site and the rest of the cube comes into play (depicted using doubled dashed lines in Fig.\ \ref{fig:FP}(c)).
These three Kitaev interactions effectively induce a Heisenberg coupling of the emerging spin $\vec S_e$ with 
$\vec S_1$.
Therefore a second Kondo effects gets activated resulting in a further screening of an effective spin-$1/2$ degree of freedom, and thus $S^\imp$ decreases from $\log 2$ to zero at a crossover temperature $T^*$.
This can be further understood by examining the low-energy spectrum of the Kitaev-cube,
that will be discussed in detail in Sec.\ \ref{subsec:spectrum}.
$S^\imp$ for a standard Kondo model consisting of a spin-1/2 impurity and using the same value of Kondo coupling $J$ is also plotted in Fig.\ \ref{fig:cube_entropy}(a) using dashed lines.
For $K'=0$, the impurity entropy is just a sum of the impurity entropy of the Kondo model and of the 7-site cluster with one side removed (hence an additional entropy of $\log 2$ is added to it for comparison).
Numerically, the two Kondo temperatures do, however, not match which is a numerical artifact.
%and the quantitative effect due to this will be explained in Sec.\ \ref{subsec:spectrum}.

Figure\ \ref{fig:cube_entropy}(b) plots $S^\imp$ for various values of the Kondo coupling $J$ for a fixed value of the Kitaev couplings $K'=5$ x $10^{-4}$.
$T^*$ is defined to be the temperature at which $S^\imp$ crosses $\frac{1}{2}\log 2$ while decreasing from $\log 2$ to zero.
Note that a very small value of $K'$ is chosen deliberately in order to avoid the trivial decoupled fixed point behavior that occurs for $K'>T_K$.
The nature of the curves are qualitatively similar to that in Fig.\ \ref{fig:cube_entropy}(a).
The crossover temperature $T^*$ decreases as $J$ is increased (or as $T_K$ increases).
It should however be noted that the above discussed two-stage screening can only be achieved for a very small range of the Kondo coupling $ 0.29 \lesssim J \lesssim 0.35$.
This is because if $J \gtrsim 0.35$, the Kondo temperature shifts to higher values and it is challenging to distinguish $T_K$ separately from the decrease in the entropy of the Kitaev cube by itself.
On the other hand, for $J \lesssim 0.28$, the Kondo temperature becomes smaller than K'.

Figure \ref{fig:tstar} illustrates how the crossover temperature $T^*$ depends on the different parameters of the model.
%We define $T^*$ to be the temperature at which the impurity entropy $S^\imp$ crosses $\frac{1}{2}\log 2$ .
Figure \ref{fig:tstar}(a) plots $T^*$ on a logarithmic scale against $-1/K'$ for various values of $10^{-5}<K'<10^{-4}$ keeping $J=0.3$,
[same as parameters as in Fig.\ \ref{fig:cube_entropy}(a)].
The crossover temperature $T^*$ is found to vary linearly with the inverse of $1/K'$, i.e., $T^*\propto \exp(-c/K')$.
Figure\ \ref{fig:tstar}(b), plots $T^*$ on a logarithmic scale against the Kondo temperature $T_K$, that in turn depends on the Kondo coupling $J$, for various values of $0.29\leq J \leq 0.35$ keeping $K'=5$x$10^{-4}$ constant [same parameters as in Fig.\ \ref{fig:cube_entropy}(b)].
As is evident from the linear nature of the plot, $T^*$ is found to be proportional to $\exp(-T_K)$.
Fig.\ \ref{fig:tstar}(c) combines both sets of data $T^*$ is plotted against $-T_K/K'$.
In addition to the sets of data in sub-figures (a) and (b), we have also plotted similar sets of data obtained for various values of $K'$ keeping $J=0.32$ fixed.
All the sets of data are found to collapse on the same straight line.
Hence the crossover temperature behaves as
\begin{align}
T^*\sim \exp(-\eta T_K/K'),
\end{align}
where $\eta=\eta_c$ (for the Kitaev-cube) is a constant that depends solely on the geometry of the impurity cluster.

\subsubsection{Kitaev Tetrahedron}

\begin{figure}
\centering
\includegraphics[width=\columnwidth]{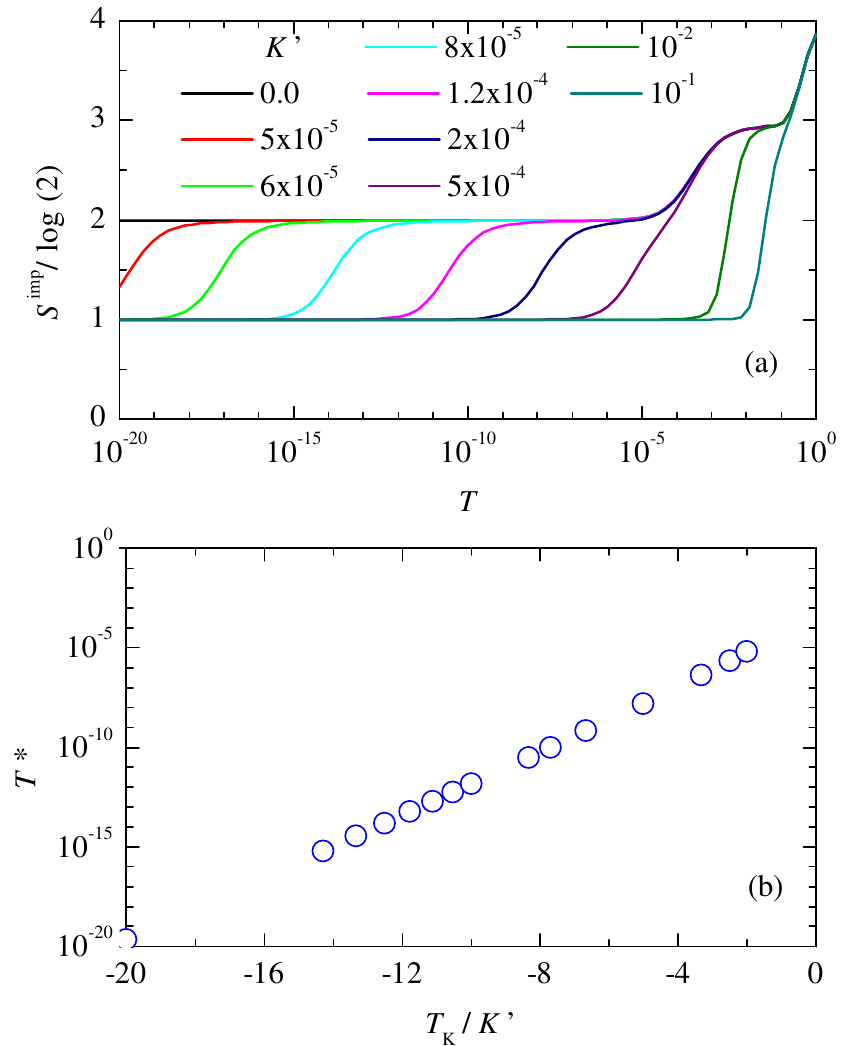}
\caption{
(a) Impurity contribution to the entropy $S_\imp$ plotted against temperature $T$ for a Kitaev-tetrahedron with coupling strength $K=0.5$ for a representative case of Kondo coupling $J=0.3$.
The plots are for different values of coupling $K'$ (refer to legend).
(b) Crossover temperatures $T^*$ extracted from the data in (a) plotted on a logarithmic scale against the ratio of $T_K/K'$.
}
\label{fig:tetrahedral_entropy}
\end{figure}

One can also perform a similar analysis for a system where the impurity consists of a Kitaev-tetrahedron as shown in Fig.\ \ref{fig:model}(b) where one corner of the tetrahedron (site 1) is Kondo-coupled to the fermionic bath.
Figure \ref{fig:tetrahedral_entropy}(a)	plots the entropy $S^\imp$ for such a model against temperature $T$ on a logarithmic scale for model parameters $K=0.5$, $J=0.3$ and for various values of $K'$ as shown in the legend.
On comparison with Fig.\ \ref{fig:cube_entropy}(a), it is found that the impurity entropy for this model shows qualitatively similar behavior as that of the cube, except that there is a residual entropy of $\log 2$ even after the two-stage screening process.
In this case, $S^\imp$ initially decreases from the high temperature value of $\log 2^4$ to $\log 8$ as the temperature is decreased.
Below the characteristic Kondo temperature $T_K$, the fermionic bath forms a Kondo singlet with the spin at site 1, decreasing the entropy by $\log 2$.
The rest of the impurity consists of a triangle with $x$, $y$, and $z$ interactions along the three bonds.
A direct diagonalization of the triangular cluster confirms that the spectrum consists of two symmetric energy levels, where each level is four-fold degenerate, thus resulting in an entropy contribution of $\log 4$.
This intermediate fixed point can also be understood in terms of a Majorana fermion formalism as in the case of the Kitaev-cube.
As the temperature is further reduced, the $K'$ interactions felicitate in the screening of an additional spin degree of freedom and hence $S^\imp$ decreases from $\log 4$ to $\log2$ (compared to $\log 2$ to 0 for the cube).
Similar to the case of the cube, the crossover temperature $T^*$  is defined to be the temperature at which $S^\imp$ crosses $\frac{3}{2} \log 2$.
Figure\ \ref{fig:tetrahedral_entropy}(b) plots $T^*$ against the ratio of $-T_K/K'$ for the plots in (a).
Once again, we find that the plot is linear implying that $T^*\propto \exp(-\eta_t T_K/K')$,
where $\eta_t$ is a constant for the tetrahedral geometry.

The residual two-fold degeneracy of the ground state arises from the two possible values of the conserved flux $W_1=\pm i$ on the face opposite to the Kondo coupled spin. This flux breaks time-reversal symmetry. This leads to two questions: (i) Why does the breaking of time reversal symmetry not affect the Kondo effect? And (ii), what will happen when the flux is promoted to a dynamical degree of freedom, e.g., by switching on a Heisenberg coupling within the cluster? These questions will be addressed in Sec.\ \ref{subsec:spectrum}.
%------------------------------------------------------------------------------------------------------

\subsection{Mapping to a two-impurity Kondo model: Numerics}
\label{subsec:2Imp}

\begin{figure}
\centering
\includegraphics[width=\columnwidth]{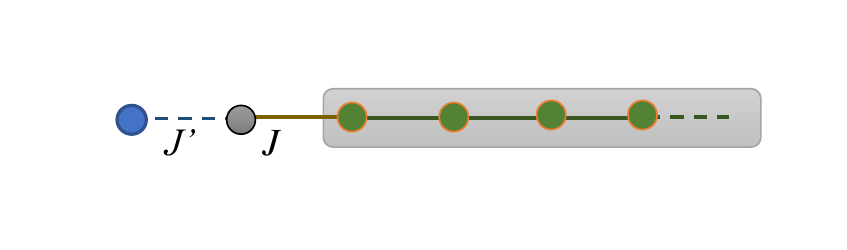}
\caption{
Schematic representation of the two-impurity Kondo model. The impurity consists of two spin-$1/2$s with an inter-impurity Heisenberg coupling $J'$ (dashed line).
One of impurity sites is coupled to the fermionic bath via Kondo coupling $J$ (solid line).
\label{fig:2I_model}
}
\end{figure}

In this section, we demonstrate that the low-temperature properties of the impurity entropy 
of the Kitaev clusters can in general be understood in terms of a simple two-impurity setup.
It was shown in the previous section that when $K' \ll T_K$, the system undergoes a two-stage Kondo screening for both geometries of the cluster considered in this study.
At first, the spin at site 1 of the impurity cluster is Kondo screened by the fermionic bath at a temperature $T_K$.
At this intermediate stage, the model consists of a Kondo singlet and a decoupled impurity consisting of one site less, with three dangling majoranas.
As the temperature is further lowered, the $K'$ terms mediate $x$, $y$ and $z$ interactions along the three bonds connecting the nearest neighbor sites.
The combined effect of these interactions is such that the rest of the impurity behaves as a spin degree of freedom connected to the rest of the system.
Thus, the cubic cluster can be mapped, at low temperatures, to a much simpler impurity consisting of just two spins-1/2's as shown schematically in Fig.\ \ref{fig:2I_model}.
Note that for the Kitaev-tetrahedron, there is an additional free $\log 2$ degree of freedom giving rise to a residual entropy, and thus the mapping to a two-impurity model is not clear.
Nonetheless, we show that these three models exhibit similar fixed points and the crossover temperature scales have similar dependencies on the model parameters.

The two-impurity Kondo model has been studied in the past comprehensibly, mostly in connection to a setup using quantum dots
\cite{Jayaprakash:81, Jones:1987,Jones:1988, Izumida:2000, Vojta:2002, Galkin:2004, Chang:2009, Bork:2011, Spinelli:2015}.
Almost all the studies considered a geometry where both of the impurities are coupled to one or two fermionic bath, in addition to being connected to each other.
In these systems, the competition between Kondo-screening and inter-impurity singlet leads to a quantum phase transition in the presence of a certain particle-hole symmetry \cite{Jones:1988, Vojta:2002}.
However, the geometry we consider in this study is different from that of previous works and is much simpler.
The model consists of two spin-1/2 impurities with an inter-impurity Heisenberg coupling $J'\vec{S}_1 \cdot \vec{S}_2$. Only one of the impurities ($\mathrm{S}_1$) is attached to a fermionic bath via Kondo exchange coupling $J$.
We demonstrate that this model exhibits thermodynamic properties that have similar temperature dependencies as that of both the Kitaev-cube and tetrahedron, except for the residual $\log 2$ entropy in the case of the latter.

\begin{figure}
\centering
\includegraphics[width=\columnwidth]{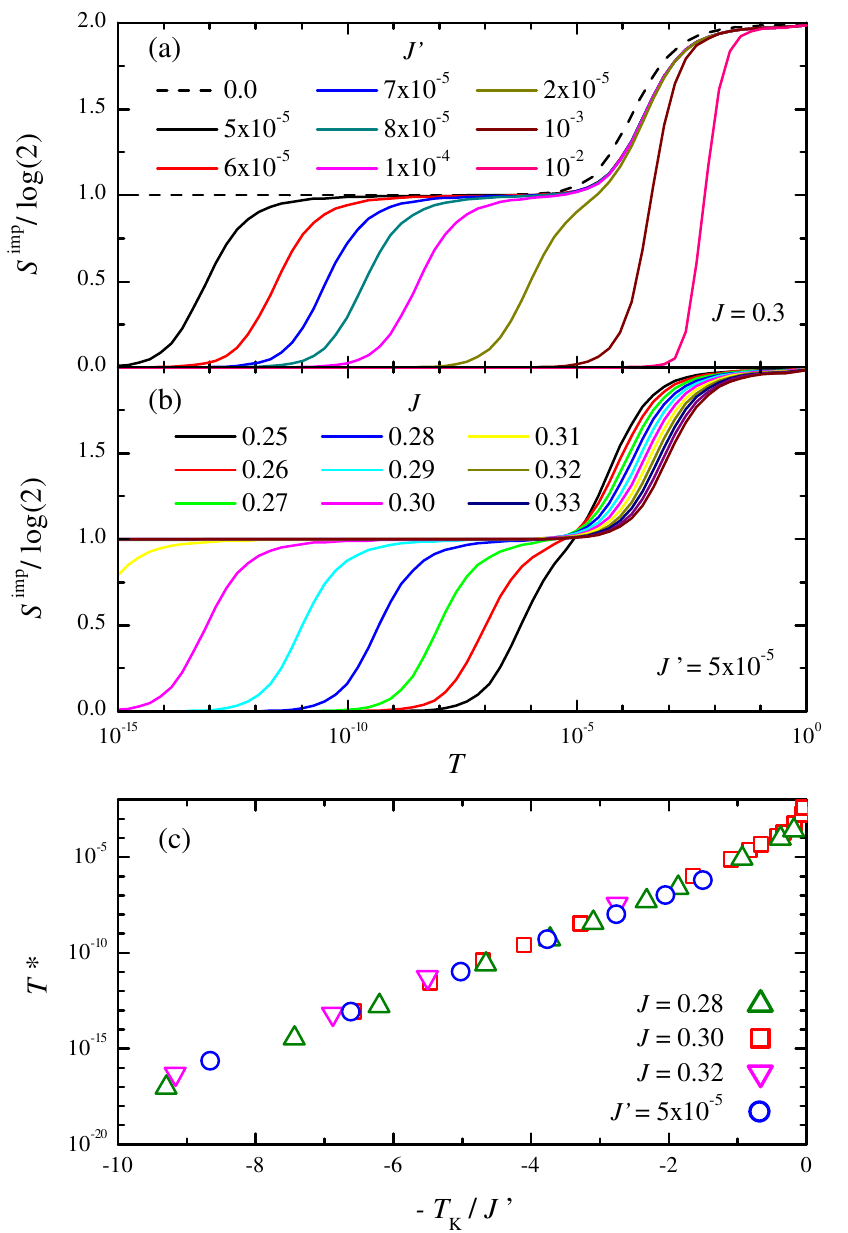}
\caption{
(a) Impurity contribution to the entropy $S^\imp$ plotted against temperature $T$ for a two-impurity model as shown in Fig.\ \ref{fig:2I_model} for a representative case of Kondo coupling $J=0.3$ and for various values of inter-impurity coupling $J'$ (refer to legend).
(b) $S_\imp$ plotted against temperature $T$ for various values of Kondo coupling $J$, keeping $J'=5\times 10^{-5}$ fixed. 
(c) Crossover temperature $T^*$ plotted on a logarithmic scale against $-T_K/J'$ for the plots in (a) and (b)
as well as for cases where $J'$ is varied keeping $J=0.28$ and $0.32$ fixed, respectively.
}
\label{fig:2I_S}
\end{figure}

Figure\ \ref{fig:2I_S}(a) plots the impurity contribution to the entropy $S^\imp$ against temperature $T$ (plotted on a logarithmic scale) for the two-impurity model for a fixed value of Kondo coupling $J=0.3$ and varying the inter-impurity coupling $J'$.
The behavior is similar to that of the Kitaev-clusters, as there is a competition between the Kondo and Heisenberg inter-impurity couplings.
For $J' > T_K$, the two impurity spins forms a singlet, and hence $S^\imp$ decreases from the high-temperature value of $\log 4$ to zero at a temperature $T\sim J'$.
However, for $J' < T_K$, $\mathrm{S}_1$ is screened by the bath at the corresponding Kondo temperature $T_K \approx 10^{-4}$, thereby reducing $S^\imp$ from $\log 4$ to $\log 2$. 
As the temperature is further decreased, $S^\imp$ decreases from $\log 2$ to 0 at a crossover temperature $T^*$ that varies as $\log T^* \propto -1/J'$.
%Similar to that of the cubic cluster, we define the Kondo temperature $T_K$ to be the temperature at which $S^\imp$ crosses $\frac{3}{2}\log 2$ and the crossover temperature $T^*$ to be the temperature at which $S^\imp$ crosses $\frac{1}{2}\log 2$.  
Figure\ \ref{fig:2I_S}(b) plots $S^\imp$ for various values of the Kondo coupling $J$ keeping $J'=5\times10^{-5}$ fixed instead.
The nature of the graph for  low temperatures is once again similar to that of the cubic cluster shown in Fig.\ \ref{fig:cube_entropy}(b).
$S^\imp$ decreases from $\log 4$ to $\log 2$ at $T=T_K(J)$, and then decreases further from $\log 2$ to 0 at the crossover temperature that varies as $\log T^* \propto -T_K$.
The combined dependence of $T^*$ on both $J'$ and $T_K$ (and indirectly its dependence on $J$) is shown in 
Fig.\ \ref{fig:2I_S}(c), that plots $T^*$ on a logarithmic scale against the ratio $-T_K/J'$.
Data are shown for the two plots in (a) and (b), using hollow squares and circles respectively.
We have also added similar sets of data for cases where $J'$ is varied while keeping the Kondo coupling fixed at $J=0.28$ and $J=0.32$ respectively.
All the sets of data are found to be linear and collapse onto each other, thus confirming that
the crossover temperature can be expressed as
$T^* \sim \exp(-\eta T_K/J')$, where $\eta=\eta_\mathrm{2IK}$ (for the two-impurity geometry) is a constant coefficient.

The two-stage screening effect can be explained as follows:
In the limit that $J'<T_K$, the inter-impurity coupling is effectively zero at high temperatures.
The spin at site-1 is Kondo-screened by the electrons in the fermionic bath at the characteristic Kondo temperature
$T_K\approx \tilde{D}\exp(-1/\rho_0J)$, ignoring higher order corrections,
where $\rho_0$ is the density of states of the bath electrons and $\tilde D$ is some renormalized value of the bandwidth or cutoff energy of the bath electrons.
After the formation of the Kondo singlet, the $\vec S_1$ becomes a part of the fermionic bath via repeated spin-flip scatterings (more generally speaking, the impurity spin is absorbed in the Fermi liquid comprising the bath electrons). 
The impurity density of states exhibits Abrikosov-Suhl resonance at the Fermi energy and below $T_K$, where the width of the resonance is given by the Kondo temperature $T_K$.
Thus for temperatures much lower than the Kondo temperature, $T_K$ behaves as the effective bandwidth as observed by the second spin-site $S_2$, hence $\tilde \rho_0 \sim 1/T_K$.
$S_2$ is Kondo-screened at a further lower temperature given by
$T^*\approx \tilde D \exp(-1/\tilde \rho_0J') \approx \tilde T_K \exp(-\eta T_K/J')$, where $\eta$ is an inverse proportionality constant connecting $\tilde \rho_0$ and $T_K$.
The prefactor $\tilde T_K$ is some renormalized value of the Kondo temperature (or the effective bandwidth) that adds to logarithmic corrections to the data collapse shown in Fig.\ \ref{fig:2I_S}(c) as well as those in Fig.\ \ref{fig:cube_entropy}(c).
It should be noted that the crossover temperature $T^*$ is difficult to extract even numerically.
The Kondo temperature by itself is logarithmically small energy scale.
$T^*$ has a similar expression as that of the Kondo temperature, where the logarithmic argument is itself a function of $T_K$, thus giving rise to extremely small energy scales. 

%------------------------------------------------------------------------------------------------

\subsection{Mapping to two-impurity Kondo model: Analytics}
\label{subsec:spectrum}

Why the seemingly complex Kitaev clusters behave like a simple two-spin impurity is an intriguing question.
More precisely, we need to understand why after Kondo screening of site one of the cluster, the rest of the cluster or the cluster minus a vacancy behaves at the lowest energy scales as an effective spin 1/2.
To do this, it is imperative that we examine the low-energy spectrum of the Kitaev-clusters.
This can easily be done using direct diagonalization in both the spin and Majorana representations as discussed in Sec.\ \ref{subsec:ED}.
However, one must note that in the spin representation, the eigensates of the Hamiltonian are not necessarily the eigenstates of the plaquette flux operators in the case of degenerate energy levels.
Hence, one cannot label the eigenstates using flux quantum numbers.
This problem is solved using a technique as follows.
We consider instead a Hamiltonian $H'=H+\sum_i \alpha_i \hat{W}_i$, 
where $H$ is the original Hamiltonian, $\alpha_i'\ll K$ ($i=1,..,5$) are negligible random numbers, and $\hat{W}_i$ are the six flux operators.
Thus, by adding small incommensurable perturbations that are proportional to the flux operators, one can force the eigenstates of $H'$ to also be the eigenstates of the plaquette flux operators.
(Since $\hat W_6=\hat{W}_1 \hat{W}_2 \hat{W}_3 \hat{W}_4 \hat{W}_5$, it is automatically ensured that the resulting states are also eigenstates of the $\hat W_6$ along with the rest of the flux operators.)

\subsubsection{Kitaev-Cube}
\label{subsubsec:cube}

\begin{figure}
\centering
\includegraphics[width=\columnwidth]{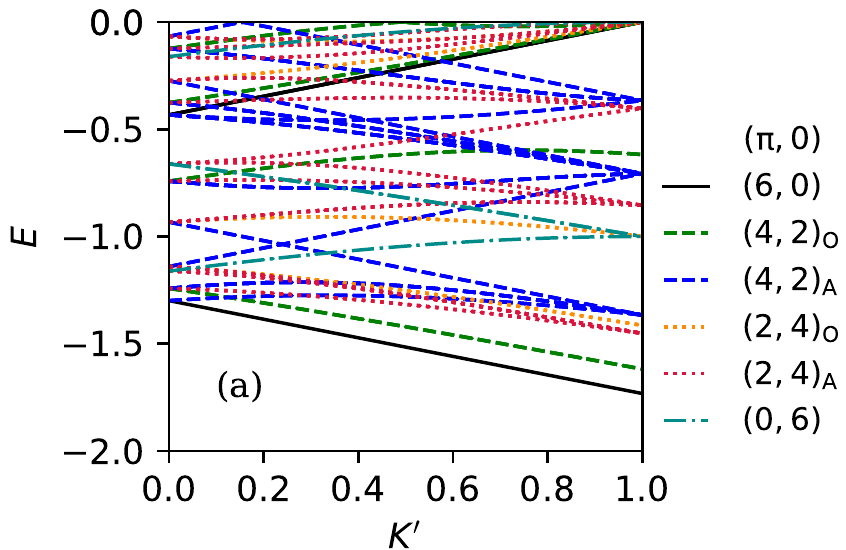} \\
\includegraphics[width=\columnwidth]{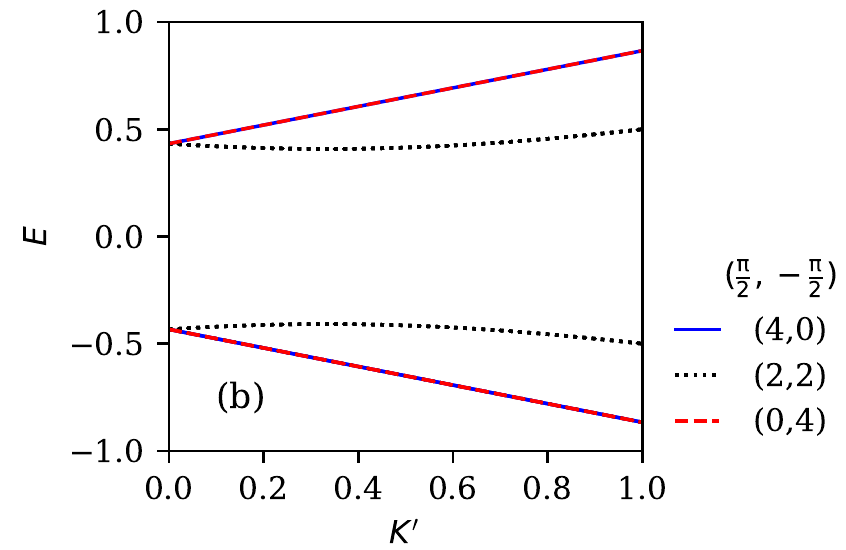}\\
\includegraphics[width=\columnwidth]{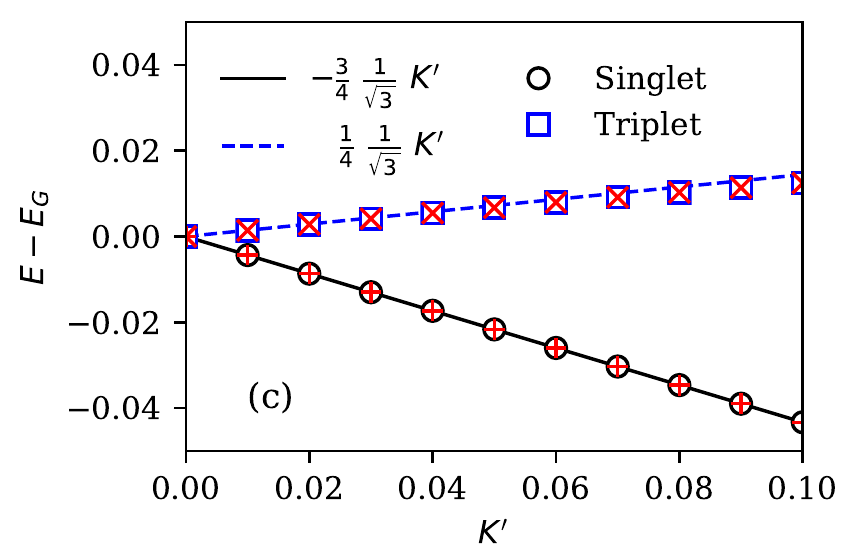}
\caption{
(a) Energy spectrum of the Kitaev-cube shown in Fig.\ \ref{fig:model}(a) with Kitaev-couplings $K=1.0$.
The couplings $K'$ between site 1 and the neighboring three sites are varied from 0.0 to 1.0.
The states are distinguished based on the number of $\pi$ and $0$ fluxes through the sides of the cube.
(b) Energy spectrum of a Kitaev-tetrahedron  shown in Fig.\ \ref{fig:model}(b) with Kitaev-couplings $K=1.0$.
The states are distinguished based on the number of $\pm \pi/2$ fluxes through the sides of the tetrahedron.
(c) A magnified plot of the ground state flux configurations and the lowest lying three-fold degenerate excited states after subtracting of the ground state energy for $K'=0$.
The data for the cube are plotted with hollow symbols whereas those for the tetrahedron are plotted using (+) and ($\times$) for the ``singlet" and ``triplet" states respectively.
The solid (dashed) line shows the corresponding singlet (triplet) energy level for two spin-$1/2$'s connected via Heisenberg coupling with strength $K'/\sqrt 3$.
}
\label{fig:spectrum}
\end{figure}

\begin{figure}
	\centering
	\includegraphics[width=\columnwidth]{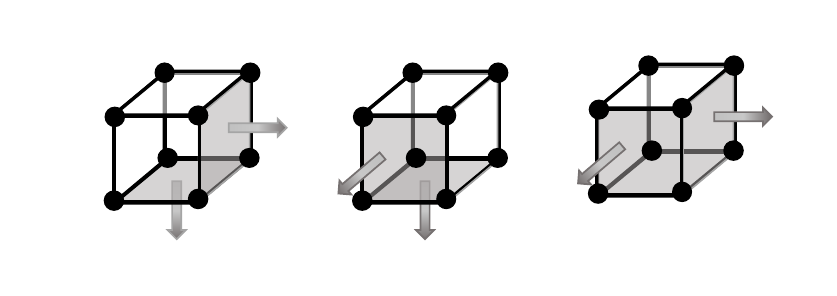}
	\caption{
		Flux configurations of the three degenerate states that effectively behave as a triplet.
		In each of the three configurations, there is a zero flux associated with the shaded pair of adjacent sides (with arrows pointing outwards). The rest of the four sides have a $\pi$ flux associated with them.
	}
	\label{fig:fluxtr}
\end{figure}

We consider a cubic cluster where the Kitaev couplings $K=1$ (on an arbitrary scale) and the three couplings $K'$ (connecting site 1 with the three nearest neighbors) are varied from 0 to 1.
Figure\ \ref{fig:spectrum}(a) plots all the energies for $E<0$.
The spectrum is symmetric with respect to zero, so the positive energies can be easily visualized from the negative spectrum.
Since the plaquette length equals 4, the flux through each plaquette is either $\pi$ or $0$.
Thus, the eigenstates can be labeled by counting the number of $\pi$ and $0$ fluxes through the six sides of the cube.
We use a notation to label each state by a pair of numbers of the form: 
$(n_\pi,n_0)$ where $n_\pi$ ($n_0$) denotes the number of $\pi$ (0) fluxes, satisfying $n_\pi+n_0=6$.
For example, the cube has a unique ground state for all values of $0 < K'\leq 1$ with $\pi$ fluxes through all the 6 sides. This state is plotted with a solid black line and labeled as $(6,0)$ in the legend accompanying the figure.
Since flux excitations must occur in pairs, the only other possible flux configurations are $(4,2)$, $(2,4)$ and $(0,6)$, plotted using dashed, dotted and dash-dotted lines respectively.
In addition to the above classification, for the eigenstates containing two 0 or $\pi$ fluxes, one can note whether they occur in opposite or adjacent sides.
This is also distinguished using a subscript of `O' and `A' respectively in the legend. 

The information that is most relevant to the results in our case is contained in the lowest lying eigenstates, i.e., the ground and the first excited states of the system, since the effect of higher excited states are negligible at low temperatures.
As mentioned above, the ground state is found to be non-degenerate for $K'>0$ and has $\pi$ fluxes through all the 6 plaquettes.
The first excited states are found to be three-fold degenerate and contain two $0$ fluxes through a pair of adjacent plaquettes.
The three possible flux configurations are illustrated in Fig.\ \ref{fig:fluxtr} where the sides associated with $\pi$ fluxes are unmarked.
In contrast, the pair of sides associated with a $0$ flux are shaded and shown with arrows coming out of the surface of the cube.
The ground state and the three excited states behave like a singlet and triplet states respectively, thus
justifying the mapping to a two-impurity system as discussed in the previous section.
This is further illustrated in Fig.\ \ref{fig:spectrum}(c) that plots (using open symbols) the pair of states on a magnified scale where $0\leq K' \leq 0.1$.
The ground state energy $E_{G,0}$ for the cluster with $K'=0$ is subtracted from all energy levels.
It can also be shown that for $K' \ll K$, the energies of ``singlet" and ``triplet" states can be approximated by $ -3K'/4\sqrt{3}$ and $K'/4\sqrt{3}$ respectively
(see Appendix \ref{appendix:pert_analytics}).
These are plotted using solid and dashed lines in Fig.\ \ref{fig:spectrum}(c) and agrees with the numerical data.
Thus the lowest lying energy levels resemble that of a Heisenberg interaction between the two spin 1/2's, $\vec S_1$ and $\vec S_e$, with an effective coupling strength $K'/\sqrt{3}$.

We find that the emerging spin  $\vec S_e=\frac{1}{2} \vec \sigma_e$ can, for example, be written as
\begin{eqnarray}
\begin{split}
\sigma_e^x &=& \sigma_4^x \sigma_8^y \sigma_7^z \sigma_6^z \sigma_5^y 
&=& -i b_4^y u_{48} u_{87} u_{76} u_{65} b_5^z,  \\
\sigma_e^y &=& \sigma_5^x \sigma_8^z  \sigma_7^z  \sigma_6^x  \sigma_2^y 
&=& -i b_5^z u_{58} u_{87} u_{76} u_{62} b_2^x,  \\
\sigma_e^z &=& \sigma_2^y \sigma_6^y \sigma_5^z \sigma_8^x \sigma_4^x 
&=& -i b_2^x u_{26} u_{65} u_{58} u_{84} b_4^y. \\
\end{split}
\label{emergentSpin}
\end{eqnarray}
This formula can be checked by projecting the operators onto the two-dimensional Hilbert space spanned by the two-degenerate ground state wave-functions of the cube with one site ($\vec S_1$) removed.
In Eq.\ \eqref{emergentSpin}, we have written the spin operators also in terms of the dangling Majorana fermion and a string of gauge links connecting the two.
Here, the 'dangling Majoranas' refer to the three single dots shown in Fig.\ \ref{fig:FP},
and the presence of a gauge string ensures that $\sigma_e$ is a gauge-invariant operator in the physical Hilbert space.
Note that alternative representations of $\sigma_e$ also exist
(e.g., $\tilde \sigma_e^x = \sigma_4^z \sigma_3^y \sigma_7^y \sigma_8^z \sigma_5^x  =
-i b_4^y u_{43} u_{37} u_{78} u_{85} b_5^z$),
where the gauge string is along a different path.
Using the ground-state flux configuration, one can show that these operators are identical in the low-energy sector (but differ at higher energies).
In ground-state flux sector, one can choose a gauge such that the product of the $u_{ij}$'s in Eq.~\eqref{emergentSpin} equals 1 for all three spin operators.
In this case, the emergent spin is simply given by the dangling Majorana fermions
 \begin{eqnarray}
\sigma_e^x = -i b_4^y b_5^z, \quad \sigma_e^y = -i b_5^z b_2^x , \quad \sigma_e^z = -i b_2^x b_4^y
\label{emergentSpinDangling}
\end{eqnarray}

According to our analysis, one can therefore view the two-stage Kondo effect for small $K'$ in at least two different ways.
One point of view is to argue that the Kondo coupling induces strong quantum fluctuations in the three flux configurations of Fig.\ \ref{fig:fluxtr} defining the triplet state.
For sufficiently strong Kondo coupling, the triplet combines with the singlet to form two separate spins,
which are highly entangled states of strongly fluctuating flux- and Majorana configurations.  
An arguably simple picture emerges when one does not track the dynamics of the flux configuration but instead describes the cube with one spin removed simply by the three dangling Majorana states of Eq.\ \eqref{emergentSpinDangling} (or, by the equivalent, gauge invariant formula of Eq.\ \eqref{emergentSpin}). While the emergent spin is a highly non-local object, it nevertheless allows for a straightforward mapping to a two-spin Kondo model.

\subsubsection{Kitaev-Tetrahedron}
\label{subsubsec:tetra}
A similar analysis can be performed for the Kitaev-tetrahedron.
The energy spectrum for such a calculation is plotted in Fig.\ \ref{fig:spectrum}(b).
Since the bond length of the plaquette operators is three, it is known that each side of the tetrahedron can have a plaquette of $\pm \pi/2$.
We use the same notation as in the previous subsection and label each state using a pair of numbers that denote the number of $\pi/2$ and $-\pi/2$ fluxes respectively through the sides, the possible configurations being (4,0), (2,2) and (0,4).
The spectrum appears to be simpler than that of the cube, with only four energy levels that are symmetric with respect to zero.
However, one added complexity is that the ground state is doubly degenerate consisting of either $\pi/2$ or $-\pi/2$ flux through all the four sides for all vales of $K'>0$.
The first excited states consist of six degenerate levels that can be classified as follows:
For the ground state flux configuration with $\pi/2$ ($-\pi/2$) fluxes through all the four sides, there are three flux excitations with $-\pi/2$ ($\pi/2$) fluxes through two of the three sides adjacent to site 1.
For small values of $0<K'<0.1$, the energy levels (after subtracting off the energy for $K'=0$) are also plotted in Fig.\ \ref{fig:spectrum}(c) using symbols ($+$) and ($\times$) for the ground states and excited states respectively.
Surprisingly, the data for the tetrahedron lies on top of those of the cube implying that in the limiting case, they have the same dependence on $K'$.
Thus the ground state of each flux configuration and the three corresponding excited states behave like ``singlet" and ``triplet" states with an effective Heisenberg coupling of $K'/\sqrt{3}$ (see Appendix \ref{appendix:pert_analytics}.

We can again provide an explicit construction of the emergent spin $\vec S_e$ which is responsible for the two-stage Kondo effect for small $K'$.
In the presence of the Kondo coupling, the fluxes in contact with $\vec S_1$ become fluctuating quantum variables, but the flux $\hat W_1 = -i \sigma_2^x \sigma_3^z \sigma_4^y$ on the opposite face of the tetrahedron remains conserved.
In the flux sector with flux $-\pi/2$ or $\hat W_1=-i$, the emergent spin is given by
\begin{eqnarray}
\begin{split}
\sigma_e^x &=& \sigma_4^y \sigma_3^z &=& i b_4^z u_{43} b_3^y, \\
\sigma_e^y &=& \sigma_2^z \sigma_4^x &=& i b_4^z u_{42} b_2^x,\\
\sigma_e^z &=& \sigma_3^x \sigma_2^y &=& i b_2^x u_{23} b_3^y.
\end{split}
\label{emergentSpinTetra}
\end{eqnarray}
For these spin-operators, one finds that
$\sigma_e^\alpha \sigma_e^\beta = i \epsilon_{\alpha \beta \gamma} \sigma_e^\gamma (i{\hat{W}}_1)$, i.e., one recovers the well-known commutation relations only in the flux sector where $\hat W_1=-i$. 
The presence of the static flux $\hat W_1$ implies that time reversal symmetry is broken as the expectation value $\langle \sigma_2^x \sigma_3^z \sigma_4^x \rangle$  is finite. Why does the breaking of time-reversal symmetry not destroy the Kondo effect and is this an artifact of the Kitaev limit or is it valid on more general grounds (e.g., when an extra Heisenberg coupling is added)?
To understand the nature of the ground state degeneracy and role of the broken time reversal symmetry it is useful to analyze the symmetries of the tetrahedron.
% which is done in Appendix XXX.
Here the important symmetries are the 180 degrees rotations of the spins around either the $x-$, $y-$ or $z-$axis.
A rotation around the $z$ axis is, for example, described by $S_x \rightarrow -S_x$, $S_y \rightarrow-S_y$, and $S_z \rightarrow S_z$. Remarkably, these rotations leave the flux invariant.
The three spin-rotation symmetries ensure that despite the broken time-reversal symmetry, no magnetic field emerges.
This ensures that a Kondo effect can be realized even in a system where time-reversal symmetry is broken.

\begin{figure}
	\centering
	\includegraphics[width=\columnwidth]{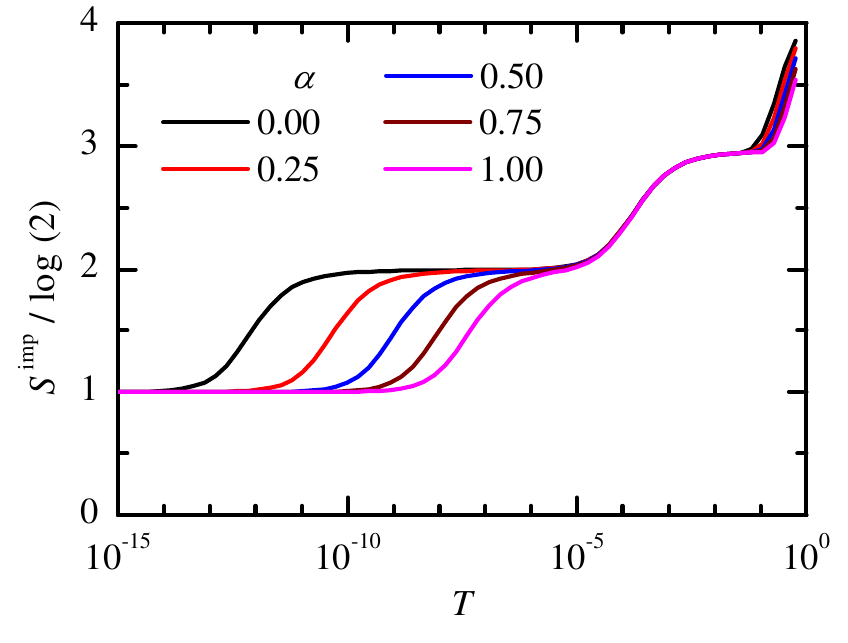}
	\caption{
		Impurity contribution to the entropy $S_\imp$ plotted against temperature $T$ for a Kitaev-tetrahedron with Kitaev couplings $K=0.5$ and $K'=10^{-4}$, and Kondo coupling $J=0.3$.
		Additional Heisenberg couplings ($J_H$ and $J^{\prime}_H$) between the sites of the cluster are also introduced such that $J_H=\alpha K$ and $J^\prime_H = \alpha K'$.
		The plots are for different values of $\alpha$ (refer to legend).}
	\label{fig:tetrahedral_entropy_Heisenberg}
\end{figure}

If we add a nearest-neighbor Heisenberg coupling to the tetrahedron Hamiltonian, the flux $\hat W_1$ is not conserved anymore. Nevertheless, the ground-state of the cluster remains two-fold degenerate. The degeneracy of the (non-Kramers) doublet in the absence of the Kondo coupling is protected by a combination of time-reversal symmetry and rotation symmetries (the degeneracy can, e.g., be lifted if the coupling on one leg of the tetrahedron is changed so that  rotation symmetries are lost). The Kondo coupling induces quantum fluctuations of the non-Kramers doublet and therefore the question emerges whether the combination of Kondo and Heisenberg coupling leads to a screening of the remaining flux degree of freedom. In Fig.~\ref{fig:tetrahedral_entropy_Heisenberg} we show the impurity entropy $S^{\imp}$ against temperature for the Kitaev-tetrahedron for Kitaev couplings $K=0.5$ and $K'=10^{-4}$, and Kondo coupling $J=0.3$, Additional Heisenberg couplings $J_H$ and $J^{\prime}_H$ are also added along the bonds (where similar to the notations adopted for the Kitaev couplings, $J_H^\prime$ denotes the coupling between $\vec S_1$ and the other three spins).
Plots are for different values of the proportionality constant $\alpha $ (see legend) where
$J_H=\alpha K$ and $J^\prime_H=\alpha K^\prime$.
The numerical result clearly shows that the residual entropy remains at $\log 2$: while the non-Kramers doublet does become a dynamical degree of freedom due to the combined presence of Kondo and Heisenberg couplings, the doublet remains unscreened.
Technically, this is a consequence of the fact that the coupling of the conduction electrons to the doublet is an irrelevant operator. Time-reversal and rotation symmetries forbid a linear coupling of the doublet operator $\vec \tau$ with the conduction electron spin. Only irrelevant higher order terms involving the product of two electron spin operators are symmetry allowed. We have checked this statement by calculating the spectrum of a Kitaev-Heisenberg tetrahedron in the presence of a magnetic field $\vec B$ applied to site $1$ of the cluster. We find that the splitting of the ground state is proportional to ${\vec B}^2-B_x B_y-B_y B_z-B_y B_x$ and therefore quadratic in the field.

\begin{figure}
	\includegraphics[width=\columnwidth]{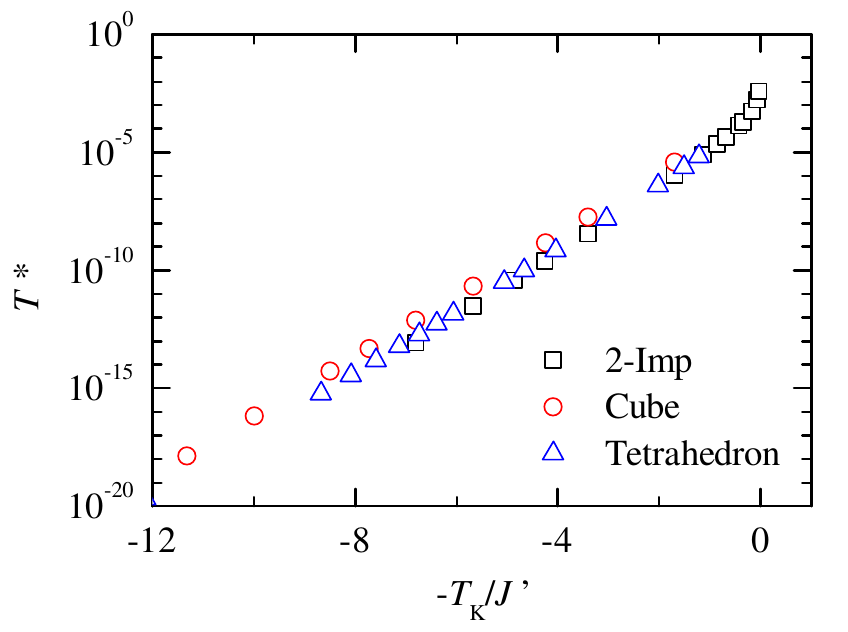}
	\caption{Combined plot of crossover temperature $T^*$ against the ratio of the coupling $T_K/J'$ for cases where the impurity consists of Kitaev clusters (cube and tetrahedron) and two-spin $1/2$s connected with Heisenberg coupling $J'$. For the Kitaev-clusters, the effective coupling $J'$ equals $K'/\sqrt{3}$.
	}
	\label{fig:comparison}
\end{figure}

Thus, we have demonstrated qualitatively as well as quantitatively that upon tuning the couplings $K'$ connecting one corner of a Kitaev-cube to its three nearest neighbors, the lowest-level energy states behave like a set of ``singlet-triplet" similar to that of two spin-1/2s connected by a Heisenberg coupling, with the effective Heisenberg coupling given by $J'=K'/\sqrt{3}$.
Now let us conclude this discussion by combining  the results for the Kitaev-cube as well as the two-impurity Kondo model.
Figure \ref{fig:comparison} plots the crossover temperature $T^*$ for both the Kitaev-cube as well as the -tetrahedron along with that for an impurity consisting of two spin-1/2s connected via Heisenberg coupling $J'$.
The $x$-axis is chosen to be the ratio $-T_K/J'$,
where the effective coupling $J'=K'/\sqrt{3}$ for the Kitaev clusters.
As is seen from the plot, the three sets of data are parallel to and lie on top of each other thus confirming the simple mapping of the Kitaev-clusters to that of the two-impurity model at low energy scales.
The minor difference in the intercept arising from different pre-factors in the exponential relations is possibly due to truncation errors and an incorrect estimation of the Kondo temperature.

%--------------------------------------------------------------------------------------------------

\subsection{Plaquette Fluxes}
\label{subsec:PF}

\begin{figure}
\centering
\includegraphics[width=\columnwidth]{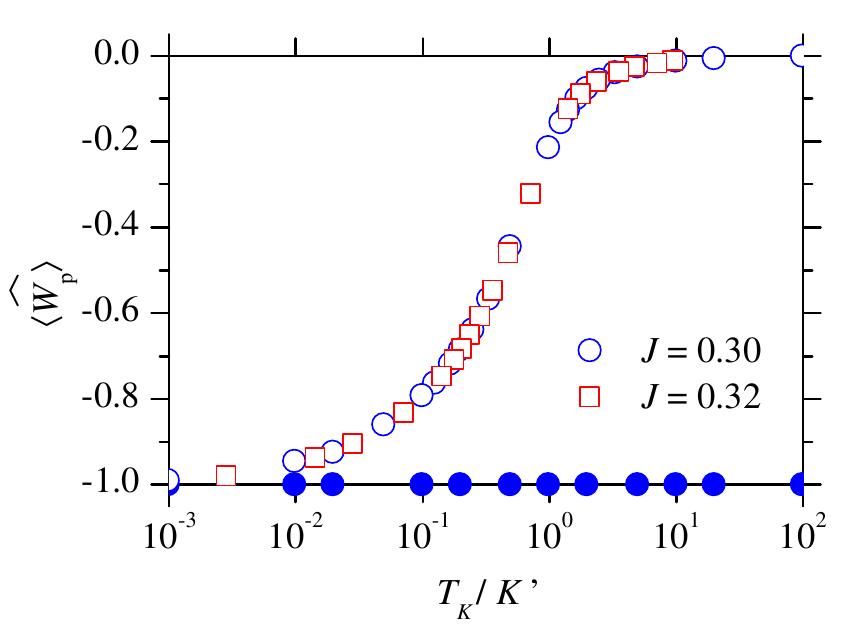}
\caption{
	Expectation values of the flux operators $\hat W_p$ for the Kitaev-cube plotted against $-T_K/K'$ on a log scale for J=0.3 and J=0.32 (see legend). The Kitaev couplings $K=0.5$.
	The plots are for $\hat W_p = \hat W_\text{Adj.} = \hat W_{1,2,3}$ (open symbols) and for $\hat W_p = \hat W_\text{Opp.} = \hat W_{4,5,6}$ (solid symbols).
	$\hat W_\text{Opp.}$ commutes with the Hamiltonian and has an expectation value of -1 ($\pi$ flux) for all values of $T_K/K'$,
	whereas $\hat W_\text{Adj.}$ do not commute and the expectation value decreases from 0 (Kondo limit) 
	to -1 (Kitaev limit) as $K'$ is increased.
}
\label{fig:flux}
\end{figure}

In Sec.\ \ref{subsec:PF}, we have introduced six plaquette flux operators $W_p$ ($p=1,2,..,6$) corresponding to the six sides of the Kitaev cube.
In the presence of a non-zero coupling of site 1 of the cube to a fermionic bath, the flux operators corresponding to the sides adjacent to site 1, i.e., $\hat W_\text{Adj.}=\hat W_{1,2,3}$ do not commute with the Hamiltonian whereas the flux operators corresponding to the sides that are opposite to site 1, i.e., $\hat W_\text{Opp.}=\hat W_{4,5,6}$ still commute with the Hamiltonian.
This is demonstrated in Fig.\ \ref{fig:flux} that plots the expectation values of the plaquette flux operators $\langle \hat W_p\rangle$ at zero temperature for various values of Kitaev coupling $K'$, against the ration $T_K/K'$.
The plots are for a fixed value of $K=0.5$, and for two values of the Kondo coupling, $J=0.3$ and $J=0.32$.
$\langle \hat W_\text{Adj.} \rangle$ and $\langle \hat W_\text{Opp.} \rangle $ are shown using hollow and solid symbols respectively.
For large values of $K'\gg T_K$ (Kitaev limit), all the plaquette operators $\hat W_p$ have an expectation value of -1 implying $\pi$ flux through all the plaquettes.
However as $K'$ is decreased, $\hat W_\text{Adj.}$ averages out and hence increases from $-1$ at $K'\sim T_K$ and approaches 0 as $K'\rightarrow0$.
Thus for large Kondo coupling, the fluxes exhibit strong fluctuations.
%Note that in this limit, $\hat W_\text{Adj.} \neq 1/3$ due to cancellation from all the four flux configurations i,e. the ``singlet" ground states and the three ``triplet" states of the Kitaev-cube impurity.
Also, the data collapse onto each other signifying a single universal function that solely depends on $T_K/K'$. 
On the other hand, the expectation values of $\hat W_\text{Opp.}$ is pinned at $-1$ as expected.

As discussed in Sec.\ \ref{subsec:PF}, the impurity flux defined by $\hat W_I$ = $\hat W_1 \hat W_2 \hat W_3$ is still a good quantum number for $J\geq 0$.
In the limit $J=0$, all the flux operators are good quantum numbers and hence the ground state of the system can also be labelled in terms of the eigenvalues $w_p$ of all the plaquette flux operators $\hat W_p$.
In this limit, for the ground state, $w_p=-1$ for all values of $p=1,2,..,6$ implying that there is $\pi$ flux through all the sides.
Hence the corresponding eigenvalue of the impurity flux operator $\hat W_I$ also equals -1.
We found that even away from this limit, i.e.,  for values of $J>0$, $\langle \hat W_I \rangle =-1 $.
Thus no flux transition is observed in the Kitaev-cube impurity systems.
It is imperative to note the differences between this result and those obtained in previous studies that focused mainly on the effect of defects on a two-dimensional Kitaev-honeycomb lattice.
The defects can be either in the form of vacancies \cite{Willans:2010,Willans:2011} on the lattice or can be due to the presence of a magnetic impurity Kondo coupled to one site in the lattice \cite{Vojta:2016, Dhochak:2010,Das:2016}
The two dimensional Kitaev-honeycomb lattice has been shown to capture an impurity $\pi$ flux at a vacancy site.
Thus upon coupling a magnetic impurity to a site and then tuning the Kondo coupling between the impurity and the lattice site, a flux transition from impurity flux was observed as the Kondo coupling was tuned from zero (Kitaev limit) to $J>>K$ (vacancy limit).
This difference can be attributed to the following comments:
(i) Although the two sets of problems are similar and try to investigate the physics arising in due to the competition of both Kitaev and Kondo physics, yet the exact geometries are vastly different and that affects specific properties of the system,
(ii) It has been shown that for a finite size two-dimensional Kitaev honeycomb lattice consisting of just three plaquettes with open boundaries, upon attaching a Kondo impurity to the center, there exists a flux transition from $0$ to $\pi$ flux.
Thus the finite size of the clusters is unlikely to be the reason for the absence of a flux-transition.
(ii) It has been shown that for three dimensional Kitaev materials, a vacancy does not bind a flux \cite{Sreejith:2016}.
Hence it is likely that due to the closed geometry of the clusters considered in this study, that do not exhibit a flux transition inspite of the demonstrating interesting effects due to the competition of Kitaev and Kondo couplings.

For the case of the Kitaev-tetrahedron, the situation is quite different.
In the limit of $J=0$, the ground state is doubly degenerate, with the plaquette fluxes of either $w_p=i$ ($\pi/2$ flux) or $w_p=-i$ ($-\pi/2$ flux) through all the sides.
Thus if one calculates the expectation value $\langle \hat W_p \rangle$ of the individual plaquettes, they average out to zero.
This is true even for cases where $J>0$.
Hence it is not possible to extract any useful information from a plot of the $\langle \hat W_p \rangle$ for the Kitaev-tetrahedron.
Due to the presence of the time reversal symmetry, the expectation values of the fluxes are all zero.

% ==============================================

\section{Discussion}
\label{sec:discussion}

To summarize, we have introduced a novel approach to investigate the physics of Kitaev materials by looking at finite-size Kitaev-clusters in the context of quantum impurities.
The clusters by themselves are interesting constructs that can help in the understanding of Kitaev physics.
Due to the finite size of clusters, it is relatively simple to the find out the spectrum using exact diagonalization techniques.
We have studied models where these Kitaev-clusters, specifically the case of cube and tetrahedron, is coupled to a bath of non-interacting fermions.
There exists a competition between the Kitaev couplings $K'$ and the Kondo temperature $T_K$.
For the case where $K^\prime > T_K$, the cluster decouples from the bath,
whereas for the more interesting case where $K^\prime < T_K$ , the model undergoes a two-stage screening process.
By studying the spectrum of the finite clusters, we were able to map the set of models to that exhibited by a model comprising of two spin-1/2's interacting with a fermionic bath, where the emergent spin $1/2$ consists of a highly entangled state of both Majorana and flux degrees of freedom.
For the Kitaev-cube, we also showed the effect of couplings on the plaquette fluxes, particularly the strong fluctuations of the flux degrees of freedom in the Kondo limit (small values of Kitaev couplings $K'$).

One interesting aspect of the results is the fact that after the Kondo screening of the one site of the Kitaev-cluster,
the interaction between the screened site and rest of the cubic cluster can be described by that of two spin-1/2's interacting via Heisenberg interactions.
Thus a vacancy created at the corner of the cube due to Kondo-screening generates an effective local moment degree of freedom, which can be viewed as arising from dangling Majorana bonds.
This is similar to what has been obtained in studies on vacancies in the Kitaev-honeycomb lattice, i.e., the formation of a local moment at the site adjacent to the vacancies that has a non-trivial dependencies on applied magnetic field \cite{Willans:2010,Willans:2011}.
Although there have been studies exploring the effect of bond-disorder in the Kitaev-honeycomb lattice in a fixed gauge sector \cite{Knolle:2019}, an investigation on the effect of a finite concentration of vacancies is lacking.
In particular, it would be interesting to understand how an increase in the concentration of the lattice vacancies effect (or destroys) spin-fractionalization and the ordering of $\mathbb{Z}_2$ gauge fluxes at low temperatures.

For the Kiteav-tetrahedron, a residual flux degree of freedom leads to a two-fold degeneracy of the ground-state and a $\log 2$ residual impurity entropy of the Kitaev-Kondo model. We have shown that this residual entropy is not quenched when the flux becomes a dynamical degree of freedom due to an extra Heisenberg coupling.
The non-Kramers doublet is protected by spin-rotation symmetries and time reversal and does not undergo a Kondo effect.
Our results show that even in the presence of extra Heisenberg coupling terms, the fractionalization of spins into Majorana and gauge degrees of freedom provides a useful language to describe the effective low-energy theory governing the physics of our clusters.
For the future, it will be interesting to connect the physics of the impurity models to properties of bulk Kitaev Kondo models using, for example, ideas from dynamical mean field theory.

\acknowledgments
We would like to thank Simon Trebst and Tim Eschmann for useful discussions.
%This work was supported by the DFG within the CRC 1238 Projects C03 (T.C and R.B) and C02 (A.R).
This work was supported by the DeutscheForschungsgemeinschaft (DFG, German Research Foundation) -  Projektnummer 277146847 - CRC 1238 (Projects C02 and C03).

\appendix

\section{Numerical RG}
\label{appendix:NRG}

The Numerical Renormalization Group (NRG) was originally developed to explain the properties of magnetic impurities in a metallic bulk, the continuous band of conduction electrons is mapped to a semi-infinite tight binding chain of fermions also known as the Wilson chain.
The discretization parameter $\Lambda \geq 1$ acts as a control parameter, and the continuum limit is achieved at the limit $\Lambda \rightarrow 1$.
Previous studies on a variety of quantum impurity problems have shown that, depending on the problem,
a range of $2\leq \Lambda \leq 6$ to be practical and useful, 
although $\Lambda$ values as high as 9 have been used to study critical exponents at quantum phase transitions \cite{Ingersent:2002,Chowdhury:2015}.
The impurity is connected to the first site of the tight-binding Wilson chain and the strength of the hopping coefficients fall off as $\Lambda^{-L/2}$ with the length $L$ of the chain.
The ground state is then determined using an iterative diagonalization of the Wilson chain, adding one fermionic site at each iteration.

In this study the complex Kitaev-cluster acts as a quantum impurity that is Kondo-coupled to a semi-infinite fermionic chain.
For a cluster consisting of $N_s$ spins, the basis states at iteration zero are formed using
(i) the basis states of the Kitaev cluster using a spin representation consisting of $2^{N_s}$ states
($\mid \up\up\up...\up\rangle$, $\mid \dn\up\up...\up\rangle$, .., $\mid \dn\dn\dn...\dn\rangle$) $\otimes$
(ii) the 4 configurations of the $c_0$ site ($|0\rangle$, $\mid\up\rangle$, $\mid\dn\rangle$, and $\mid\up\dn\rangle$).
In the subsequent iterations, one fermionic site is added and the scaled Hamiltonian is iteratively diagonalized in the standard fashion.
Due to the presence of the Kitaev couplings, the $z$ component of the total spin of the system, $S^z$, is no longer conserved,
and hence one cannot label the eigenstates using the eigenvalues of the $S^z$ operator.
The total charge measured from half-filling can still be used as a good quantum number and the Hamiltonian can be block-diagonalized by breaking up the the Hilbert space into subspaces labeled by the charge quantum number.
However, the lack of spin symmetries increases the overall computational time.
In the NRG, the total number of states increases by four at each iteration and hence it becomes impractical to work with all the states.
So, the high energy excitations are truncated at the end of each iteration by either keeping only a fixed number of lowest lying states or by using an energy cutoff. 
However, for the Kitaev cube for example, the number of basis states at iteration zero equals $2^8 \times 4 = 1024$ (compared to $2 \times 4 = 8$ for the single impurity Kondo model and $4\times4=16$ for the single impurity Anderson model).
In order to not loose too much information at the initial iterations,
all the states are kept for the first few iterations before implementing the truncation scheme.
Overall, the large number of states at iteration zero can lead to large truncation effects at the initial iterations, and it is impractical to counterbalance this by increasing the number of kept states by a feasible number.
Although it is impossible to get rid off this problem completely, we found that it merely renormalizes the Kondo temperature.
%and does not affect the low-temperature physics of the system qualitatively.
This is evident, for example, in Fig.\ \ref{fig:cube_entropy}(a) that shows a noticeable difference in Kondo temperatures $T_K$ for (i) a single spin (shown using dashed lines) and (ii) a Kitaev-cube with $K'=0$ such that the rest of the cluster is decoupled from site $1$. 
Thus the enormous truncation effect results in an over-estimation of the Kondo temperature.
However, we emphasize that the qualitative behavior of the model at low-temperatures is unaffected, although it is crucial that we use the numerically estimated (and hence ``modified") Kondo temperature $T_K$ in our calculations.

Thermodynamic properties of the system such as entropy can also be calculated by computing the expectation value of the appropriate operators, at the end of each NRG iteration.
To determine the temperature dependencies of an observable, we associate a temperature $T\propto \Lambda^{-N/2}$, $N$ being the iteration number.
The impurity contribution to the thermodynamic properties are determined by subtracting off the contribution from the fermionic bath alone.
We use this formalism to calculate the expectation value of the plaquette fluxes through the different sides of Kitaev clusters as will be discussed shortly.

\section{Mapping to two-impurity Kondo model: Perturbative analysis}
\label{appendix:pert_analytics}
Here, we present a perturbative treatment of the finite-clusters to obtain analytical expressions of the effective Heisenberg coupling in terms of the Kitaev couplings for both the Kitaev-cube and -tetrahedron.
We consider clusters with Kitaev couplings $K^\prime$ between site 1 and the three nearest neighbor sites and couplings $K$ for the rest of the links, similar to what was done in Sec.\ \ref{subsec:spectrum}.
The energies of the cluster can be obtained by solving the respective Kitaev clusters using the Majorana fermion formalism described in Sec.\ \ref{subsubsec:ED-majorana}.
The couplings are chosen such that $K^\prime \ll K$, and we compute the energies by keeping terms upto lowest order in $K^\prime$. 
Note that in the case of the Kitaev honeycomb lattice, one usually uses a convention such that the bond operators are taken to be positive (or negative) depending on whether they point from sublattice $A$ to $B$ (or vice versa).
For the case of the finite systems considered here, the choice of the bond operators is quite arbitrary.
As mentioned in Sec.\ \ref{subsubsec:ED-majorana}, we use a convention such that $u_{ij} = 1$ if $i>j$ and $u_{ij} = -1$ otherwise.

At first, let us compute the energies of the relatively easier case of the Kitaev-tetrahedron.
The ground state is found to be double degenerate with flux configurations $(\frac{\pi}{2},\frac{\pi}{2},\frac{\pi}{2},\frac{\pi}{2})$ and $(-\frac{\pi}{2},-\frac{\pi}{2},-\frac{\pi}{2},-\frac{\pi}{2})$ (in the unprojected Hilbert space this flux configuration is over-counted by a factor of 8 reflecting 8 possible gauge choices).
As an example, let us choose a flux configuration given by the following values of $u_{ij}$:
\begin{align}
\centering
\begin{split}
u_{12}=-1,\ u_{23}=1,\ u_{13}=-1,\\
u_{14}=-1,\ u_{24}=-1,\ u_{34}=1. 
\end{split}
\end{align}
The $A$-matrix defined by Eq.\ \eqref{eq:A-matrix} for the set of bond variables is given by
\begin{align}
iA=\frac{i}{2}
\begin{pmatrix}
0  & -K' & -K' & -K' \\
K' & 0  & K  & -K \\
K' & -K & 0   & K \\
K' & K  & -K   & 0
\end{pmatrix}
.
\end{align}
The eigenvalues of $iA$ (or the single-particle energies of the fermionic spectrum) are found to be
\begin{align}
\epsilon_i= \{	\pm \frac{\sqrt{3}}{2}K', \pm \frac{\sqrt{3}}{2}K \}.
\end{align}
The ground state energy of the many-particle Majorana spectrum is then given by
\begin{align}
E_0= -\frac{1}{2} \sum_{i=1}^2 \epsilon_i = -\frac{\sqrt{3}}{4}(K+K').
\end{align}
Upon subtracting the value of $E_1(K'=0)$, we get
\begin{align}
E_0(K')-E_0(K'=0)=-\frac{\sqrt{3}}{4}K' = -\frac{3}{4} \frac{1}{\sqrt{3}}K'
\label{eq:s-tetra}
\end{align}

For the first excited state, there are six gauge-invariant states (we ignore the first excited unphysical states that arise in the Majorana spectrum). 
Each of those states has a multiplicity of 8 (arising from different gauge choices), thus the total number of degenerate levels in the unprojected Hilbert space is 48.
As an example, let us consider a set of $\{u_{ij}\}$ values as follows:
\begin{align}
u_{12}=-1,\ u_{23}=-1,\ u_{13}=1,\\
u_{14}=-1,\ u_{24}=1,\ u_{34}=-1. 
\end{align}
This corresponds to one of the pair of triplet states with a flux configuration of $(-\frac{\pi}{2},\frac{\pi}{2},-\frac{\pi}{2},\frac{\pi}{2})$.
The $A$-matrix is given by:
\begin{align}
iA=\frac{i}{2}
\begin{pmatrix}
0   & -K' &  K' & -K' \\
K'  &  0  & -K  &  K \\
-K' &  K  &  0  & -K \\
K'  &  K  &  K  & 0 \\
\end{pmatrix}
,
\end{align}
and has the following eigenvalues:
\begin{align}
\pm \frac{1}{2\sqrt{2}}\big(3K^2 + 3K'^2 \pm \sqrt{9K^4+14K^2K'^2+9K'^4}\big)^{1/2}.
\end{align}
Expanding the eigenvalues and keeping terms upto linear order in $K^\prime$, 
we obtain the single-particle energy levels:
\begin{align}
\epsilon_1 =	\frac{\sqrt{3}}{2}K \text{ and } \epsilon_2 = \frac{K'}{2\sqrt{3}}.
\end{align}
The (physical) energy of the first-excited levels is given by
\begin{align}
E_1 = -\frac{1}{2}(\epsilon_1-\epsilon_2)=\Big( -\frac{\sqrt{3}K}{4} + \frac{K'}{4\sqrt{3}} \Big).
\end{align}
Upon subtracting the energy at $K'=0$, we obtain
\begin{align}
E_1(K')-E_1(K'=0)=\frac{K'}{4\sqrt{3}}=\frac{1}{4}\frac{1}{\sqrt{3}}K'.
\label{eq:t-tetra}
\end{align}
Thus, it is evident that the ground and the first excited levels have a dependency on $K'$ similar to that of a pair of singlet-triplet states that arise for two-spins if they are mutually connected via Heisenberg interactions with an effective strength $J'=K'/\sqrt{3}$.

Let us repeat the calculations for the cube as well for the sake of completeness.
For the ground state, we choose, as an example, the following one of the $2^7$ possible configurations of $\{u_{ij}\}$ that leads to the unique ground state characterized by a gauge-independent flux configuration of $(\pi,\pi,\pi,\pi,\pi,\pi)$:
\begin{align}
\begin{split}
u_{12}=1,u_{23}=1,u_{34}=1,u_{14}=1,\\
u_{56}=-1,u_{67}=-1,u_{78}=-1,u_{58}=-1,\\
u_{15}=-1,u_{26}=-1,u_{37}=-1,u_{48}=-1.
\end{split}
\end{align}
The $A$-matrix is given by
\begin{align}
iA=\frac{i}{2}
\begin{pmatrix}
0  &  K' &  0  &  K' & -K' &  0 &  0 & 0 \\
-K' &  0  &  K  &  0  &  0  & -K &  0 & 0 \\
0  & -K  &  0  &  K  &  0  &  0 & -K & 0 \\
-K' &  0  & -K  &  0  &  0  &  0 &  0 & -K \\
K' &  0  &  0  &  0  &  0  & -K &  0 & -K \\
0  &  K  &  0  &  0  &  K &  0  & -K & 0 \\
0  &  0  &  K  &  0  &  0 &  K  &  0 & -K \\
0  &  0  &  0  &  K  &  K &  0  &  K & 0 \\
\end{pmatrix}
,
\end{align}
with eigenvalues
\begin{align}
\pm \frac{\sqrt{3}}{2}K, \pm \frac{\sqrt{3}}{2}K, \pm \frac{\sqrt{3}}{2}K, \pm \frac{\sqrt{3}}{2}K'.
\end{align}
The ground state energy of the fermionic many-body state is given by:
\begin{align}
E_0= -\frac{1}{2} \sum_{i=1}^4 \epsilon_i = -\frac{\sqrt{3}}{4}(3K+K').
\end{align}
For the first excited state, let us consider a configuration (out of $3\times 2^7$ such possibilities) given by:
\begin{align}
\begin{split}
u_{12}=1, u_{23}=1, u_{34}=1, u_{14}=1,\\
u_{56}=-1,u_{67}=-1,u_{78}=-1,u_{85}=-1,\\
u_{15}=-1,u_{26}=-1,u_{37}=-1,u_{48}=-1,
\end{split}
\end{align}
The $A$-matrix
\begin{align}
iA=\frac{i}{2}
\begin{pmatrix}
0 & -K' &  0  & -K' & -K' &  0 &  0 & 0 \\
K'&  0  &  K  &  0  &  0  & -K &  0 & 0 \\
0 & -K  &  0  &  K  &  0  &  0 & -K & 0 \\
K'&  0  &  0  &  0  &  0  &  0 &  0 & -K \\
K'&  0  &  0  &  0  &  0  & -K &  0 & -K \\
0 &  K  &  0  &  0  &  K &  0  & -K & 0 \\
0 &  0  &  K  &  0  &  0 &  K  &  0 & -K \\
0 &  0  &  0  &  K  &  K &  0  &  K & 0 \\
\end{pmatrix}
\end{align}
has eigenvalues:
\begin{align*}
\begin{split}
\pm\sqrt{3}K/2, \pm \sqrt{3}K/2, \\
\pm \frac{1}{2\sqrt{2}}\big(3K^2 + 3K'^2 \pm \sqrt{9K^4+14K^2K'^2+9K'^4}\big)^{1/2}.
\end{split}
\end{align*}
The structure of the eigenvalues is similar to that of the tetrahedron.
Keeping terms up to linear order in $K'$, the single-particle energies can be expressed as:
\begin{align}
\epsilon_1,\epsilon_2,\epsilon_3 = \frac{\sqrt{3}K}{2}, \text{ and }
\epsilon_4 = \frac{K'}{2\sqrt{3}}.
\end{align}
The (physical) first-excited state energy is given by:
\begin{align}
E_1 & = \frac{1}{2}(\epsilon_1+\epsilon_2+\epsilon_3-\epsilon_4)\\
& = -\frac{1}{2}\Big( \frac{3\sqrt{3}K}{2}-\frac{K'}{2\sqrt{3}}\Big).
\end{align}
Upon subtracting $E_0(K'=0)$ from the expressions for $E_0$ and $E_1$, we get respectively
\begin{align}
\begin{split}
E_0(K') -E_0(K'=0)&= -\frac{3}{4} \frac{1}{\sqrt{3}}K', \text{ and }\\
E_1(K')-E_1(K'=0)&= \frac{1}{4} \frac{1}{\sqrt{3}} K'.
\end{split}
\label{eq:st-cube}
\end{align}
Equation\ \ref{eq:st-cube} has the same form as those in Eqs.\ \ref{eq:s-tetra} and \ref{eq:t-tetra} respectively.
Thus, for small values of $K' \ll K$, both the Kitaev-cube and -tetrahedron can be mapped to an impurity consisting of two-spins with an effective Heisenberg coupling given by $K'/\sqrt{3}$.

\end{document}